\documentclass[aps,pra,reprint,superscriptaddress]{revtex4-2}

\usepackage{amsmath}
\usepackage{amssymb}
\usepackage{mathtools}
\usepackage{enumitem}
\usepackage{times}

\usepackage[hiresbb]{graphicx}

\newcommand{\Star}[1]{#1\ensuremath{^\ast}\kern-\scriptspace}
\newcommand{\CStar}{\Star{\ensuremath{\mathrm{C}}}}

\DeclareMathOperator*{\argmin}{arg\,min}

\newcommand{\defeq}{\coloneqq}

\DeclareMathSymbol{\rstr}{\mathclose}{AMSa}{"16}

\newcommand{\HS}[1]{\mathcal{#1}}

\newcommand{\abs}[1]{\lvert #1 \rvert}
\newcommand{\dabs}[1]{\left\lvert #1 \right\rvert}

\newcommand{\norm}[1]{\lVert #1 \rVert}
\newcommand{\lnorm}[2]{\lVert #1 \rVert_{#2}}

\newcommand{\ev}[1]{\langle #1 \rangle}

\newcommand{\lev}[2]{\langle #1 \rangle_{\hspace{-1.0pt}#2}}
\newcommand{\dlev}[2]{\left\langle #1 \right\rangle_{\!#2}}

\newcommand{\inpr}[2]{\langle #1, #2 \rangle}

\newcommand{\linpr}[3]{\langle #1, #2 \rangle_{#3}}
\newcommand{\dlinpr}[3]{\left\langle #1,\, #2 \right\rangle_{\!#3}}

\newcommand{\stdv}[1]{\sigma(#1)}
\newcommand{\lstdv}[2]{\sigma_{\hspace{-1.0pt}#2}(#1)}
\newcommand{\dlstdv}[2]{\sigma_{\hspace{-1.0pt}#2}\left(#1\right)}

\newcommand{\id}{\mathrm{Id}}

\newcommand{\dom}[1]{\mathrm{dom}(#1)}
\newcommand{\ran}[1]{\mathrm{ran}\,#1}
\newcommand{\cran}[1]{\overline{\mathrm{ran}}\,#1}

\newcommand{\gr}[1]{\mathrm{gr}(#1)}

\newcommand{\cob}[2]{R_{#2}(#1)}
\newcommand{\qob}[2]{S_{\hspace{-0.5pt}#2}(\mathcal{#1})}

\newcommand{\spb}[1]{#1^{\hspace{-0.5pt}\ast}}

\newcommand{\pb}[2]{#1_{\!#2}^{\hspace{-0.5pt}\ast}}
\newcommand{\pf}[2]{#1_{\hspace{-1pt}#2\ast}}
\newcommand{\dpb}[2]{\pb{#1}{#2}}
\newcommand{\dpf}[2]{#1_{\!#2\ast}^{\phantom{\hspace{-0.5pt}\ast}}}

\newcommand\inv[1]{#1\raisebox{1.15ex}{$\hspace{0.5pt}\scriptscriptstyle-\!1$}}
\newcommand{\invpb}[2]{\inv{(\pb{#1}{#2})}}
\newcommand{\invpf}[2]{\inv{(\dpf{#1}{#2})}}

\newcommand{\err}[3]{\varepsilon_{\hspace{-0.5pt}#3}(#1;#2)}
\newcommand{\rerr}[3]{\tilde{\varepsilon}_{\hspace{-0.5pt}#3}(#1;#2)}

\begin{document}


\title{A Universal Formulation of Uncertainty Relations for Errors and Local Representability of Quantum Observables}


\author{Jaeha Lee}
\email[]{lee@iis.u-tokyo.ac.jp}
\affiliation{Institute of Industrial Science, The University of Tokyo, Chiba 277-8574, Japan.}



\begin{abstract}
A universal formulation of uncertainty relation for quantum measurements is presented with additional focus on the representability of quantum observables by classical observables over a given state.  Owing to the simplicity and operational tangibility of the framework, the resultant general relations admit natural operational interpretations and characterisations, and are thus also experimentally verifiable.  In view of the universal formulation, Heisenberg's philosophy of the uncertainty principle is also revisited;  it is reformulated and restated as a refined no-go theorem, albeit perhaps in a weaker form than was originally intended.  In fact, the relations entail, in essence as corollaries to their special cases, several previously known relations, including most notably the Arthurs--Kelly--Goodman, Ozawa, and Watanabe--Sagawa--Ueda relations for quantum measurements.  The Schr{\"o}dinger relation (hence the standard Kennard--Robertson relation as its trivial corollary as well) is also shown to be a special case when the measurement is non-informative.
\end{abstract}


\maketitle


\section{Introduction\label{sec:intro}}

It has been a century since the discovery of quantum theory, which---together with the theory of relativity---opened the door towards modern physics, and has since revolutionised our traditional na{\"i}ve conception of nature and reality for good.  One of the cornerstones of quantum mechanics is undoubtedly the uncertainty principle, which characterises the indeterministic nature of the microscopic world in a succinct fashion.  Following Heisenberg's seminal exposition \cite{Heisenberg_1927} of the principle, the first mathematical formulation was announced by Kennard \cite{Kennard_1927} revealing the lower bound $\hbar/2$ for the product of the standard deviations of position and momentum.  This was subsequently generalised to those of arbitrary observables $A$ and $B$ by Robertson \cite{Robertson_1929} with the familiar lower bound
\begin{equation}\label{ineq:uncertainty-relation_Kennard-Robertson}
\stdv{A}\,\stdv{B} \geq \abs{ \ev{[A, B]}} /2
\end{equation}
now being dictated by the expectation value of the commutator of the observables over the quantum wave function concerned.  Owing to its mathematical clarity and simplicity, the Kennard--Robertson (KR) relation has since been widely adopted as a standard material in textbooks as a succinct expression of quantum indeterminacy.

Despite its renowned status, The Kennard--Robertson relation has been known to incorporate in their formalism little to no concept of measurement, which Heisenberg---although his own conception of uncertainty (or \lq indeterminateness\rq\ \cite{Heisenberg_1930}) is difficult to precisely confer from the fairly vague description of his writings---did originally entertain in his various examples, one of which being the famous gamma-ray microscope Gedankenexperiment.

Given the rather unsatisfying situation, there followed a number of alternative formulations of uncertainty relations involving measurement.   Conceivably the most typical approach is the adoption of the indirect measurement scheme, which explicitly incorporates the system of an auxiliary meter device in addition to the quantum system of primary interest, thereby allowing for an intuitive interpretation of an otherwise abstract mathematical formulation of error and disturbance associated with the measurement.  Some of the most renowned fruits of this approach are the Arthurs--Kelly--Goodman (AKG) relations \cite{Arthurs_1965,Arthurs_1988} and the more recent Ozawa relations \cite{Ozawa_2003, Ozawa_2004_01}, which are followed by their refinements \cite{Branciard_2013} and modifications \cite{Ozawa_2019}.  Apart form these, uncertainty relations have also been grounded from a measure-theoretic viewpoint \cite{Werner_2004, Miyadera_2008, Busch_2013} as well as within the framework of estimation theory \cite{Yuen_1973,Watanabe_2011}.

Beyond the three orthodox relations regarding quantum indeterminacy, error, and disturbance, the uncertainty principle was also found to manifest itself as various forms of incompatibilities of diverse nature.  These include, for example, the relation for `time and energy' \cite{Mandelshtam_1945, Allcock_1969_01, *Allcock_1969_02, *Allcock_1969_03, Helstrom_1976}, entropy \cite{Hirschman_1957, Beckner_1975, Birula_1975, Deutsch_1983}, conservation law \cite{Wigner_1952, Araki_1960, Yanase_1961, Ozawa_2002_01}, speed limit \cite{Fleming_1973, Aharonov_1990, Pfeifer_1993, Margolus_1998, Giovannetti_2003, Jones_2010, Pires_2016, Shiraishi_2018}, gate implementation \cite{Ozawa_2002_02, Tajima_2018}, and counterfactuality \cite{Hall_2001, Dressel_2014, Lee_2016, Pollak_2019}.

This paper presents a universal formulation of uncertainty relations along with a family of novel inequalities that mark the trade-off relation between the measurement errors of an arbitrary pair of quantum observables under arguably the most general settings;  it is established upon conceivably the simplest and most general framework of measurements of statistical nature without any additional assumptions or reference to the specific measurement models of any kind, \textit{i.e.}, the only objects required are the tangible measurement outcomes.

The aforementioned philosophy of this paper connotes that, on top of their generality, the relations presented below admit natural operational interpretations and characterisations, and are thus also experimentally verifiable;  these seemingly innocent comments---while actually being of grave importance to empirical science---do not necessarily apply to some of the alternative formulations (including Ozawa's), which generally require objects that the measurement outcomes alone cannot dictate, as have been pointed out in Refs.~\cite{Werner_2004, Koshino_2005}.

Notably, the formulation of this paper leads to several types of trade-off relations of different nature within a unified framework, thereby granting a seamless connection among the various forms of manifestation of uncertainty in quantum theory.  In this regard, the uncertainty relations presented in this paper entail, in essence as corollaries to their special cases, various known relations including the KR, AKG, Ozawa, and Watanabe--Sagawa--Ueda (WSU) \cite{Watanabe_2011} relations mentioned above.  Apart from the derivation of the KR relation, and also the outline leading to the AKG, Ozawa, and  WSU relations, further details on these topics shall be reported in the subsequent papers of the author within appropriate contexts.

This paper is organised as follows.  The first three sections following this introduction are devoted to the presentation of the universal framework.  In Section~\ref{sec:setup}, a brief introduction to the basic concepts and facts regarding quantum measurements are given.  In Section~\ref{sec:pullback_and_pushforward}, quantum measurements are then found to induce an adjoint pair of state-dependent maps, termed the pullback and pushforward.  Section~\ref{sec:partial-inverse} introduces another key concept of the universal formulation, namely the procedure of inverting a not necessarily injective closed operator, termed the standard partial inverse.

Equipped with the tools and concepts introduced so far, the subsequent three sections introduce the main results of this paper.  In Section~\ref{sec:error}, the definitions of error of quantum measurements are introduced, which are followed by useful equivalence conditions on which a measurement becomes errorless.  Section~\ref{sec:uncertainty_relation} presents a novel family of uncertainty relations regarding quantum measurements.  Refined versions of the relations are also presented in Section~\ref{sec:local_joint-describability}, in which the measurements for each of the pair of observables may be chosen differently as long as they admit a local joint-description.

The remaining two sections discuss the important consequences of the results of this paper established so far.  In Section~\ref{sec:uncertainty-principle}, the relations are found to entail an important connotation regarding the measurement of a pair of incompatible (non-commutative) quantum observables, whereby Heisenberg's philosophy of the uncertainty principle is revisited;  in view of the universal framework, it is restated as a refined no-go theorem, albeit perhaps in a weaker form than was originally introduced.  The final Section~\ref{sec:reference_to_other_relations} serves as a reference to the previous studies, in which the new relations are found to entail, in essence as a corollary to their special cases, several notable relations including the Arthurs--Kelly--Goodman, Ozawa, and Watanabe--Sagawa--Ueda relations for quantum measurements.  The Schr{\"o}dinger relation, from which the standard Kennard--Robertson relation follows as its trivial corollary as well, is also demonstrated to be a special case when the measurement is non-informative.

The final Section~\ref{sec:discussions} is devoted to discusstions.

Part of the framework and the results presented in this paper have been reported in the earlier manuscripts \cite{Lee_2020_01_preprint,Lee_2020_01_Entropy} of the authors, to which the reader is referred as appropriate.  Further details on the physical ramifications of the universal formulation as well as their mathematical descriptions shall be reported in subsequent papers of the author.

\section{Measurements\label{sec:setup}}

For the presentation of the framework, let $Z(\HS{H})$ denote the state space of a quantum system, which will be hereafter modelled as the convex set of all the density operators $\rho$ on a Hilbert space $\HS{H}$.  Its classical counterpart $W(\Omega)$ will be modelled as the convex set of all the probability distributions $p$ on a sample space $\Omega$.

\subsection{Quantum Measurement}

The primary objects of interest in this paper are the maps from quantum-state spaces to classical-state spaces that maintains the consistency with the concept of probabilistic mixture of states, namely, the affine maps $M : Z(\HS{H}) \to W(\Omega)$ that take density operators $\rho$ to probability distributions $p = M\!\rho$ satisfying the consistency condition $M(\lambda \rho_{1} + (1-\lambda)\rho_{2}) = \lambda M\!\rho_{1} + (1-\lambda) M\!\rho_{2}$ for all $\rho_{1}, \rho_{2} \in Z(\HS{H})$, $0 \leq \lambda \leq 1$.

\begin{figure}
\centering
\includegraphics[hiresbb,clip,width=0.40\textwidth,keepaspectratio,pagebox=artbox]{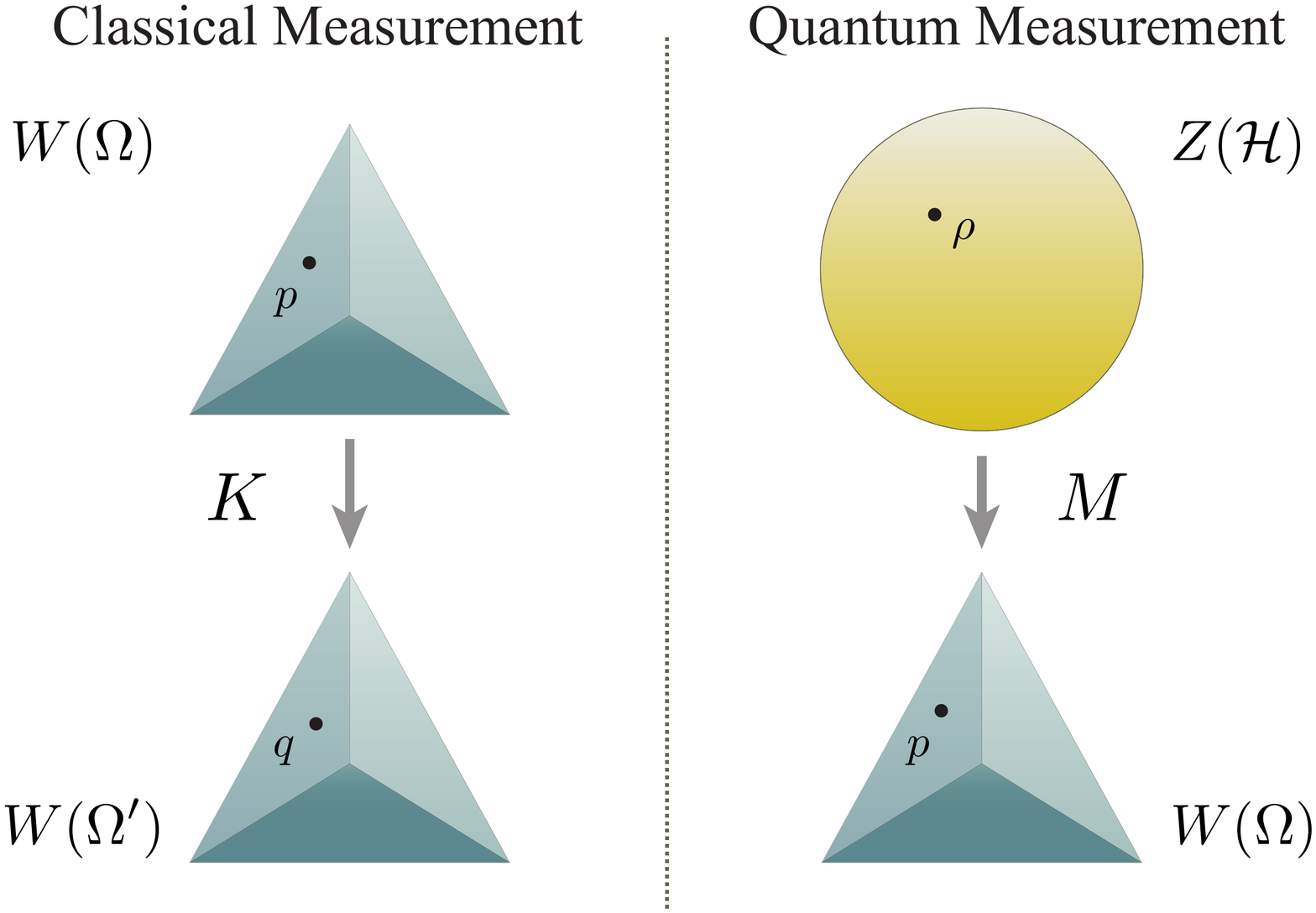}%
\caption{Classical and quantum measurements.  The space of quantum states $Z(\HS{H})$ and that of probability distributions $W(\Omega)$ are respectively depicted as a ball and a tetrahedron.  (Left) A classical measurement $K$ of statistical nature is understood to be an affine map $K : W(\Omega) \to W(\Omega^{\prime})$ that outputs a probability distribution $q = Kp \in W(\Omega^{\prime})$ on the set $\Omega^{\prime}$ of all possible outcomes. The profile of the distribution of the outputs depends on the state $p \in W(\Omega)$ of the system over which the measurement is performed.  (Right) In the same manner, an affine map $M : Z(\HS{H}) \to W(\Omega)$ is interpreted as a quantum measurement, which---dependent on the quantum state $\rho \in Z(\HS{H})$ over which the measurement is performed---generates a probability distribution $p = M\!\rho \in W(\Omega)$ on the set $\Omega$ of all possible outcomes.}
\label{fig:quantum_and_classical_measurements}
\end{figure}

The map $M$ can be interpreted in a variety of manners, \textit{e.g.}, embedding of a quantum system into a classical model, but for the purpose of this paper, let it be understood as a quantum measurement (see FIG.~\ref{fig:quantum_and_classical_measurements}).  One should indeed find this interpretation reasonable after recalling the archetypal projection measurement $M$ associated with the `measurement observable' $\hat{M}$, where the use of the diacritic over the symbol is intended to discern a Hilbert-space operator from other objects that may potentially lead to confusion.  More explicitly, let it be temporarily assumed for the sake of explanation that the observable $\hat{M}$ is a non-degenerate Hermitian matrix acting on a finite $N$-dimensional space $\HS{H}$.  Its spectral decomposition $\hat{M} = \sum_{i=1}^{N} m_{i}\, |m_{i}\rangle\langle m_{i}|$ then induces a natural affine map
\begin{equation}\label{def:projection-measurement}
M : \rho \mapsto (M\!\rho)(m_{i}) \defeq \mathrm{Tr}\bigl[ \vert m_{i}\rangle\langle m_{i} \vert \rho \bigr]
\end{equation}
that takes a density operator $\rho$ to the probability distribution $p(m_{i}) = (M\!\rho)(m_{i})$ defined on the spectrum $\Omega = \{ m_{1}, \dots, m_{N} \}$ of $\hat{M}$, which is a closed subset of the real line $\mathbb{R}$ consisting of all its eigenvalues.  The Born rule \cite{Born_1926} endows the distributions $M\!\rho$ with the standard interpretation as those describing the probability of finding each of the outcomes $m_{i} \in \Omega$ in the measurement.

Throughout this paper, the reader may simply think of the affine map $M$ as that pertaining to the familiar projection measurement \eqref{def:projection-measurement} without missing much of the essence of the subject.  Still, it is to be emphasised that the map $M$ is by no means restricted to that particular class;  indeed, if one adopts Kolmogorov's measure-theoretic formalism \cite{Kolmogorov_1933} to model probability, the measurement $M$ admits a familiar representation \cite{Holevo_2001} in terms of POVMs (Positive Operator-Valued Measures), whereby each probability measure $M\!\rho$ is essentially given in the form of \eqref{def:projection-measurement} with the elements of the POVM taking the place of the orthogonal projections.  In this paper, special attention is given to this most common and standard formalism of modelling probability, thereby making comments on subjects and facts that are measure-theory specific along the discourse as appropriate.

\subsection{Adjoint of Measurement}

An important observation is that a quantum measurement $M$, which has been introduced above as an affine map that takes a quantum state on $\HS{H}$ to a classical state on $\Omega$, induces a natural map $M^{\prime}$, termed its \emph{adjoint}, which in turn takes a classical function on $\Omega$ to a Hilbert-space operator on $\HS{H}$.  This dual notion of a quantum measurement is uniquely characterised by the relation
\begin{equation}\label{char:adjoint_quantum-measurement}
\dlev{M^{\prime}f}{\rho} = \dlev{f}{M\!\rho},
\end{equation}
which holds for all complex functions $f$ on $\Omega$ and quantum states $\rho$ on $\HS{H}$.  Here, the shorthand $\lev{X}{\rho} \defeq \mathrm{Tr}[X\rho]$ is defined for a pair of a Hilbert-space operator $X$ and a density operator $\rho$ on $\HS{H}$, whereas $\lev{f}{p} \defeq \int_{\Omega} f(\omega) p(\omega)\, d\omega$ is defined for a pair of a complex function $f$ and a probability distribution $p$ on $\Omega$.  In this regard, the projection measurement \eqref{def:projection-measurement} yet again offers a prime example;  its adjoint reads
\begin{equation}\label{def:adjoint_projection-measurement}
M^{\prime}f \defeq \sum_{i=1}^{N} f(m_{i})\, \vert m_{i}\rangle\langle m_{i} \vert,
\end{equation}
fulfilling the relation \eqref{char:adjoint_quantum-measurement}.  Under the measure-theoretic formalism of probability, the adjoint of $M$ can be represented as a map that takes the functions to their formal integrals with respect to the POVM, which, in essence, can be understood as an extension of \eqref{def:adjoint_projection-measurement} with the orthogonal projections being replaced by the elements of the POVM.

\section{Pullback and Pushforward of Quantum Measurement\label{sec:pullback_and_pushforward}}

The key to the new formulation of this paper is the observation that a quantum measurement induces an adjoint pair of \emph{local} (\textit{i.e.}, state-dependent) maps between the quantum- and classical-observable spaces (see FIG.~\ref{fig:pullback_pushforward}).  The pair, termed the pullback and pushforward of a quantum measurement, serves as an important constituent of the framework.

\subsection{The Space of Localised Observables}

The space of quantum observables will be modelled by the linear space $S(\HS{H})$ of all the self-adjoint operators on a Hilbert space $\HS{H}$.  Here, each quantum state $\rho \in Z(\HS{H})$ defines a seminorm $\lnorm{ A }{\rho} \defeq \sqrt{\lev{ A^{\dagger} A }{\rho}}$, $A \in S(\HS{H})$ on the space, thereby inducing a natural equivalence relation $A \sim_{\rho} B \iff \lnorm{ A - B }{\rho} = 0$ on it.  This allows for the classification of all the quantum observables into their equivalence classes $[A]_{\rho} \defeq \{ B \in S(\HS{H}) : A \sim_{\rho} B \}$, which collectively constitute a quotient space $S(\HS{H})/{\sim_{\rho}}$, the unique completion
\begin{equation}\label{def:loc_quantum_observables}
\qob{H}{\rho}
	\defeq \overline{ S(\HS{H})/{\sim_{\rho}} }
\end{equation}
of which shall be addressed in this paper as the space of (localised) quantum observables over $\rho$.  In a parallel manner, a probability distribution $p \in W(\Omega)$ induces a seminorm $\lVert f \rVert_{p} \defeq \sqrt{\lev{ f^{\dagger}f }{p}}$ on the linear space $R(\Omega)$ of all the real functions $f$ defined on the sample space $\Omega$.  The identification $f \sim_{p} g \iff \lVert f - g \rVert_{p} = 0$ results in the classification of the real functions into their equivalence classes $[f]_{p} \defeq \{ g \in R(\Omega) : f \sim_{p} g \}$, which collectively make up the quotient space $R(\Omega)/{\sim_{p}}$, further leading to its unique completion
\begin{equation}\label{def:loc_classical_observables}
\cob{\Omega}{p}
	\defeq \overline{ R(\Omega)/{\sim_{p}} }
\end{equation}
termed the space of (localised) classical observables over $p$ in this paper.  Here, the adjoint $A^{\dagger}$ of a Hilbert-space operator and the complex conjugate $f^{\dagger}$ of a complex function are introduced to expose the structure of the seminorms so that they respectively admit obvious extensions beyond self-adjoint operators and real functions.  As commonly practiced, with a slight abuse of notation, the equivalence classes will be denoted by one of their representatives hereafter.

\subsection{Pullback of a Quantum Measurement}

Given a quantum measurement $M : Z(\HS{H}) \to W(\Omega)$, a crucial observation regarding its adjoint is the validity of the inequality
\begin{equation}\label{ineq:non-expansiveness_adjoint}
\left\lVert f \right\rVert_{M\!\rho} \geq \left\lVert M^{\prime} f \right\rVert_{\rho}
\end{equation}
for any quantum state $\rho$ on $\HS{H}$ and complex function $f$ on $\Omega$.   A simple way to understand this would be by means of the Kadison--Schwarz inequality \cite{Kadison_1952}, which is, in a certain sense, a generalisation of the prestigious Cauchy--Schwarz inequality to \CStar-algebras.  Indeed, its straightforward application to the adjoint $M^{\prime}$ results in the evaluation $M^{\prime}(f^{\dagger}f) \geq (M^{\prime}f)^{\dagger}(M^{\prime}f)$ valid for any complex function $f$, which, combined with the characterisation \eqref{char:adjoint_quantum-measurement} of the adjoint, leads to the desired result \eqref{ineq:non-expansiveness_adjoint} as an immediate corollary.  Specifically, if one adopts the measure-theoretic formalism to model probability, thereby allowing for the representation of the quantum measurement $M$ by POVMs, the author remarks that Naimark's dilation theorem \cite{Naimark_1940, Naimark_1943} becomes directly applicable to provide a tailored and more concrete proof of the inequality \eqref{ineq:non-expansiveness_adjoint}.

The inequality \eqref{ineq:non-expansiveness_adjoint} also admits an operational and physically intuitive interpretation.  To see this, first observe its equivalence to the condition $\sigma_{M\!\rho}(f) \geq \sigma_{\rho}(M^{\prime}f)$ with the standard deviation $\lstdv{A}{\rho} \defeq (\lnorm{ A }{\rho}^{2} - \lev{ A }{\rho}^{2} )^{1/2}$ of a quantum observable $A$ over the state $\rho$ and that $\lstdv{f}{p} \defeq (\lnorm{ f }{p}^{2} - \lev{ f }{p}^{2} )^{1/2}$ of a classical observable $f$ over the distribution $p$ being introduced.  This allows one to construe the inequality \eqref{ineq:non-expansiveness_adjoint} as a statement regarding the lower bound of the efficiency of quantum measurements:  the operational cost of acquiring the expectation value of a quantum observable through measurements can never break the quantum limit imposed by the said observable.

For illustration, suppose one is to acquire the expectation value of a quantum observable $A$ through a measurement $M$.  Here, the probability distribution $M\!\rho$ of the measurement outcomes is expected to contain some information of the target quantum system, which could therefore be exploited to obtain
the desired expectation value $\lev{ A }{\rho}$ from the operationally obtained quantity $\lev{ f }{M\!\rho}$ regarding some real function $f$.  Specifically, one finds from the characterisation \eqref{char:adjoint_quantum-measurement} of the adjoint that the condition $A = M^{\prime}f$ is necessary and sufficient for the acquisition $\lev{ A }{\rho} = \lev{ f }{M\!\rho}$ to be successful \emph{globally} (\textit{i.e.}, over every quantum state).  Under this condition, the inequality \eqref{ineq:non-expansiveness_adjoint} imposes the lower bound $\lstdv{A}{\rho}$ on the operational cost $\sigma_{M\!\rho}(f)$ of the acquisition.

\begin{figure}
\centering
\includegraphics[hiresbb,clip,width=0.40\textwidth,keepaspectratio,pagebox=artbox]{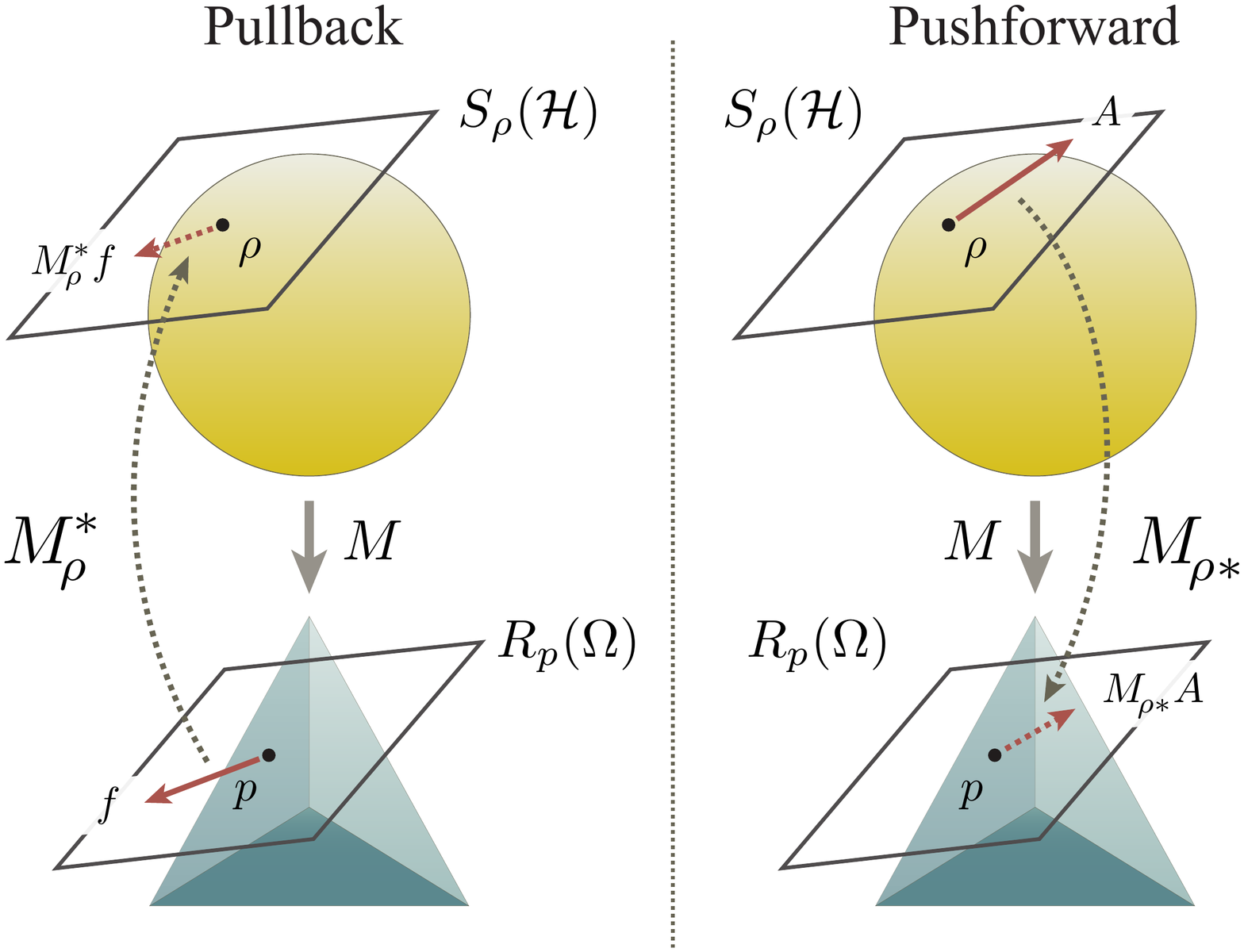}%
\caption{
The pullback and the pushforward of a quantum measurement.  (Left) A quantum measurement $M$ induces the pullback $\dpb{M}{\rho}$ from the space $R_{p}(\Omega)$ of localised classical observables to its quantum counterpart $\qob{H}{\rho}$, each of which is tangent to the respective points $p = M\!\rho \in W(\Omega)$ and $\rho \in Z(\HS{H})$ of the state spaces concerned.  (Right) The measurement $M$ equivalently generates the pushforward that maps the localised observables in the reverse direction.  The pullback and the pushforward are dual to one another in that they are characterised by the relation \eqref{char:pullback_pushforward}.  They are both non-expansive and preserve the expectation value.}
\label{fig:pullback_pushforward}
\end{figure}

A direct connotation of the inequality \eqref{ineq:non-expansiveness_adjoint} is the implication $f \sim_{M\!\rho} g \implies M^{\prime}f \sim_{\rho} M^{\prime}g$ for each quantum state $\rho \in Z(\HS{H})$.  This allows for the adjoint $M^{\prime}$, which is a map from functions to operators, to be understood as the map between the quotient spaces of real functions and self-adjoint operators;  the unique continuous extension
\begin{equation}\label{def:pullback}
\dpb{M}{\rho} : \cob{\Omega}{M\!\rho} \to \qob{H}{\rho}
\end{equation}
of the resultant map will be called the \emph{pullback of the measurement} $M$ over the quantum state $\rho$.  By construction, one finds that the pullback is non-expansive $\lnorm{ f }{M\!\rho} \geq \lnorm{ \dpb{M}{\rho} f }{\rho}$, in addition to the fact that it preserves the expectation value $\lev{ f }{M\!\rho} = \lev{ \dpb{M}{\rho} f }{\rho}$.

\subsection{Pushforward of a Quantum Measurement}

Dual to the notion of the pullback is the \textit{pushforward of the measurement} $M$.  For this, note that the norm on $\qob{H}{\rho}$ admits a unique inner product $\linpr{ A }{ B }{\rho} \defeq \lev{ \{A, B\} }{\rho}/2$ dictated by the anti-commutator $\{A,B\} \defeq AB + BA$ that reproduces the original norm $\lnorm{ A }{\rho}^{2} = \linpr{ A }{ A }{\rho}$.  The same remark also goes for the inner product $\linpr{ f }{ g }{p} \defeq \lev{ fg }{p}$ defined on $R_{p}(\Omega)$ that satisfies $\lnorm{ f }{p}^{2} = \linpr{ f }{ f }{p}$.  The pushforward 
\begin{equation}\label{def:pushforward}
\dpf{M}{\rho} : \qob{H}{\rho} \to \cob{\Omega}{M\!\rho}
\end{equation}
is then introduced as the adjoint map of the pullback \eqref{def:pullback} with respect to the above inner products;  in other words, the pushforward is characterised as the unique map that satisfies the relation
\begin{equation}\label{char:pullback_pushforward}
\dlinpr{ A }{ \dpb{M}{\rho}f }{\rho}
	= \dlinpr{ \dpf{M}{\rho} A }{ f }{M\!\rho}
\end{equation}
for all $A \in \qob{H}{\rho}$ and $f \in \cob{\Omega}{M\!\rho}$.  Owing to the non-expansiveness of the pullback, its adjoint, \textit{i.e.}, the pushforward, is also non-expansive $\lnorm{ A }{\rho} \geq \lnorm{ \pf{M}{\rho} A }{M\!\rho}$.  As with the pullback, the pushforward also preserves the expectation value $\lev{ A }{\rho} = \lev{ \pf{M}{\rho} A }{M\!\rho}$, as can be readily confirmed by choosing the constant function $f = 1$ in \eqref{char:pullback_pushforward}.

If one adopts the Kolmogorovian formalism of probability, the pushforward \eqref{def:pushforward} admits a concrete expression in terms of the familiar tools of measure and integration theory.  For this, let $\nu_{\rho}(\Delta) \defeq (\lnorm{ A + \pb{M}{\rho}\chi_{\Delta} }{\rho}^{2} - \lnorm{ A - \pb{M}{\rho}\chi_{\Delta} }{\rho}^{2})/4$ denote a signed measure defined for each measurable set $\Delta$ pertaining to a $\sigma$-algebra on the sample space $\Omega$, where $\chi_{\Delta}$ represents the measurable characteristic function (\textit{alias} indicator function) of $\Delta$.  It is then straightforward to see that $\nu_{\rho}$ is absolutely continuous with respect to the probability measure $\mu_{\rho} \defeq M\!\rho$, thereby pointing to the existence of the Radon-Nikod{\'y}m derivative \cite{Radon_1913, Nikodym_1930} of the former with respect to the latter, which in fact furnishes a useful formula
\begin{equation}\label{eq:pushforward_Radon-Nikodym}
\pf{M}{\rho} A = \frac{d\nu_{\rho}}{d\mu_{\rho}}
\end{equation}
of the pushforward;  this can be confirmed by the characterisation of the pushforward as the unique (up to the null sets of the probability measure concerned) function that fulfils the relation \eqref{char:pullback_pushforward}, which the Radon-Nikod{\'y}m derivative $d\nu_{\rho}/d\mu_{\rho}$ indeed does by construction.  For an interpretation of the pushforward in view of the formula \eqref{eq:pushforward_Radon-Nikodym}, the reader is referred to Refs.~\cite{Lee_2017, Lee_2018};  it shall also be expounded in detail in the subsequent papers of the author.

\section{Partial Inverse of the Pullback and Pushforward\label{sec:partial-inverse}}

Other important tools in the framework of the author are the `inverses' defined for the pullback and the pushforward induced by the quantum measurement, which shall be termed their (standard) partial inverses in this paper.

\subsection{Generalised Inverses}

The aim of this passage is to construct an `inverse' of a not necessarily invertible operator.  Historically, the endeavour of inverting a generally non-injective map appeared in print as early as in 1903 in the work of Fredholm \cite{Fredholm_1903}, in which he introduced, in his original terminology, a particular `pseudo-inverse' of a certain integral operator, and also perhaps implicitly in the works of Hilbert \cite{Hilbert_1904_01, Hilbert_1904_02}, in which he alluded to such an idea in 1904 in studying generalised Green functions.  Conceivably, one of the most renowned constructions in the community of physics would be the `general reciprocal' of finite-dimensional matrices, originally presented by Moore \cite{Moore_1920} in 1920, and later rediscovered by several individuals including Bjerhammar \cite{Bjerhammar_1951_01, Bjerhammar_1951_02, Bjerhammar_1958}, who recognised its role in the solution of linear systems.  Bjerhammar's results has seen further refinement and extension by Penrose \cite{Penrose_1955} in 1955, in which he specifically identified the characterisation of the reciprocal by the familiar four conditions.  Since then, the topic has received much attention and many contributions have been made.  Given the diverse terminologies for the various definitions of the `inverse', the term `generalised inverse' seems to be widely accepted as a generic word for collectively addressing the different ways of construction.

\subsection{Standard Partial Inverse of Closed Operators}

For the purpose of this paper, the author introduces below the notion of \emph{(standard) partial inverse} defined for closed linear operators, which may serve as another contribution to the study of various constructions of the generalised inverses;  in this paper, partly in order to avoid confusion with the numerous alternative constructions, the author introduces a different terminology than the preexisting terms, some of which have been reviewed above.

Recall that a linear operator $A: U \supset \dom{A} \to V$ between normed spaces $U$ and $V$ is said to be closed, if its graph
\begin{equation}\label{def:graph_of_operator}
\gr{A} \defeq \{ (x, Ax) : x \in \dom{A} \}
\end{equation}
is a topologically closed subspace of $U \oplus V$ with respect to the product topology, or equivalently, if the domain $\dom{A}$ is complete with respect to the graph norm $\lnorm{x}{A} \defeq \norm{x} + \norm{Ax}$.  Closed operators constitute a decently well-behaved class of linear maps, while encompassing a fairly wide range of important operators encountered in many practical situations;  important subclasses are those of bounded operators as well as (not-necessarily bounded) self-adjoint operators.

Let $A : \HS{H} \supset \dom{A} \to \HS{K}$ be a closed operator between Hilbert spaces $\HS{H}$ and $\HS{K}$.
Given the fact that the kernel of a closed operator is topologically closed, one readily obtains the decomposition
\begin{equation}
\dom{A}
	= \ker{A} \oplus D_{\!A}
\end{equation}
of its domain with $D_{\!A} \defeq \ker{A}^{\perp} \cap \dom{A}$.  By construction, the restriction $A \rstr_{D_{\!A}}$ of the operator $A$ to the subspace $D_{\!A}$ is trivially injective, and is also itself a closed operator;  indeed, the graph $\gr{A \rstr_{D_{\!A}}} = \gr{A} \cap \{ (x,y) \in \HS{H} \oplus \HS{K} : x \in \ker{A}^{\perp} \}$ of the restriction is topologically closed as the intersection of two closed subspaces.

Under the above settings, the author introduces the \emph{(standard) partial inverse}
\begin{equation}\label{def:partial-inverse}
\inv{A}
	\defeq \inv{\left( A \rstr_{D_{\!A}} \right)}
\end{equation}
of a closed operator $A$ as the inverse of the restriction $A \rstr_{D_{\!A}}$, which is by construction an injective operator.  Conveniently, the partial inverse of a closed operator is itself closed, as one may readily see from the fact that the graph $\gr{\inv{A}} = U^{-1} \bigl( \gr{A \rstr_{D_{\!A}}} \bigr)$ of the partial inverse and that of the restriction are characterised by one another through the unitary operator $U : \HS{H} \oplus \HS{K} \to \HS{K} \oplus \HS{H},\, (x,y) \mapsto (y,x)$.

The following properties regarding a closed operator $A$ and its partial inverse $\inv{A}$ are straightforward by construction:
\begin{align}
\inv{A}A &= P_{D_{\!A}}, \label{char:partial-inverse_01} \\
A\inv{A} &= \id_{\ran{A}}. \label{char:partial-inverse_02}
\end{align}
Here, $P_{D_{\!A}}: \ker{A} \oplus D_{\!A} \ni (x_{1},x_{2}) \mapsto x_{2} \in D_{\!A}$ is the projection onto the summand $D_{\!A}$ associated with the direct sum decomposition $\dom{A} = \ker{A} \oplus D_{\!A}$ of the domain, and $\id_{\ran{A}}$ is the identity operator defined on $\ran{A}$.  Specifically, these immediately entail the equalities
\begin{align}
A\inv{A}A &= A, \\
\inv{A}A\inv{A} &= \inv{A},
\end{align}
which are two of the familiar characterising conditions commonly adopted by many of the constructions of the generalised inverse.

\subsection{Minimality}

The partial inverse introduced in this paper also enjoys the familiar property regarding minimality that is widely found in common among many constructions of the generalised inverse.  Let $y \in \dom{\inv{A}} = \ran{A}$ be an element pertaining to the domain of the partial inverse of a closed operator $A$.  Then, among the elements $x \in \dom{A}$ for which $Ax = y$ hold, the minimum of the norm is uniquely attained by the partial inverse, \textit{i.e.},
\begin{align}\label{char:minimality}
\norm{x} \geq \norm{\inv{A}y}
\end{align}
and $\norm{x} = \norm{\inv{A}y} \implies x = \inv{A}y$; this is a straightforward consequence of the property~\eqref{char:partial-inverse_01}, which entails that any element $x \in \dom{A}$ of the domain admits a unique decomposition $x = x_{0} + \inv{A}Ax$ into the sum of the orthogonal elements $x_{0} = x - \inv{A}Ax \in \ker{A}$ and $\inv{A}Ax \in D_{\!A}$, thereby revealing
\begin{equation}
\norm{x}^{2} = \norm{x_{0}}^{2} + \norm{\inv{A}Ax}^{2} \geq \norm{\inv{A}Ax}^{2}
\end{equation}
as desired.  In other words, the partial inverse of $A$ maps an element $y \in \dom{\inv{A}}$ to the unique argument of the minimum
\begin{equation}\label{char:minimality_argmin}
\argmin_{Ax=y} \norm{x}
	= \{ \inv{A}y \}
\end{equation}
of the norm of $\HS{H}$ restricted to the preimage of $y$ under $A$.

\subsection{Adjoint}

If the closed operator $A$ happens to be densely defined, its adjoint becomes meaningful.  In such a case, one has
\begin{equation}
D_{\!A} = \cran{A^{*}} \cap \dom{A},
\end{equation}
given the basic fact from functional analysis that the orthogonal complement $\ker{A}^{\perp} = \cran{A^{*}}$ of the kernel of a densely defined closed operator $A$ is equal to the topological closure of the range of the adjoint.  It is then fairly straightforward to see that $D_{\!A}$ is dense in the closed subspace $\cran{A^{*}}$.

It now becomes tempting to consider the adjoint of the partial inverse.  One simple way to attain this is to restrict the Hilbert space encompassing the domain $\dom{\inv{A}} = \ran{A}$ of the partial inverse to its completion so that the map \eqref{def:partial-inverse_restriction} may be understood as a densely defined Hilbert-space operator in its own right:
\begin{equation}\label{def:partial-inverse_restriction}
\inv{A} : \cran{A} \supset \dom{\inv{A}} \to \HS{H}.
\end{equation}
One may then verify using the rudimentary techniques of functional analysis that the partial inverse
\begin{equation}\label{eq:partial-inverse_adjoint}
\inv{(A^{*})} = (\inv{A})^{*}
\end{equation}
of the adjoint of a closed operator coincides with the adjoint of its partial inverse.

In consideration of the fact that the products \eqref{char:partial-inverse_01} and \eqref{char:partial-inverse_02} of the operators concerned are bounded densely defined symmetric operators, their adjoints are nothing but their unique continuous extensions;  explicitly, they respectively read $(\inv{A}A)^{*} = P_{\cran{A^{*}}}$ and $(A\inv{A})^{*} = \id_{\cran{A}}$, where $P_{\cran{A^{*}}}$ is the orthogonal projection associated with the closed subspace $\cran{A^{*}}$, and $\id_{\cran{A}}$ is the identity operator on $\cran{A}$.  These facts reveal
\begin{align}
(\inv{A}A)^{**} = (\inv{A}A)^{*} &\supset \inv{A}A, \\
(A\inv{A})^{**} = (A\inv{A})^{*} &\supset A\inv{A},
\end{align}
which may be understood as a generalisation of the remaining two characterising conditions of the familiar `general reciprocal' \cite{Moore_1920, Bjerhammar_1951_01, Bjerhammar_1951_02, Bjerhammar_1958, Penrose_1955} defined for finite-dimensional matrices.

\section{Errors of Quantum Measurement\label{sec:error}}

The pullback and the pushforward, along with their partial inverses, furnish the basis of the definitions of the error of quantum measurement and its operational characterisation within the framework of this paper.

\subsection{Error of a Quantum Measurement}

Under the tools and concepts introduced so far, the author defines the \emph{error} regarding a quantum measurement $M$ of an observable $A$ over a state $\rho$ by the amount of contraction
\begin{equation}\label{def:error_quantum-measurement}
\err{A}{M}{\rho}
	\defeq \sqrt{\lnorm{A}{\rho}^{2} - \lnorm{\pf{M}{\rho}A}{M\!\rho}^{2}}
\end{equation}
induced by the pushforward.  Non-negativity $\err{A}{M}{\rho} \geq 0$ of the error follows immediately from the non-expansiveness of the pushforward.  It is also straightforward to check the absolute homogeneity $\err{tA}{M}{\rho} = \abs{ t }\, \err{A}{M}{\rho}$, $\forall t \in \mathbb{R}$, and the subadditivity $\err{A}{M}{\rho} + \err{B}{M}{\rho} \geq \err{A+B}{M}{\rho}$ of the error.  In other words, the error \eqref{def:error_quantum-measurement} furnishes a seminorm on the space $\qob{H}{\rho}$ of localised quantum observables, which could be depicted as a `tangent' space attached to the point $\rho$ over which the measurement is performed (see FIG.~\ref{fig:pullback_pushforward}).

The error \eqref{def:error_quantum-measurement} admits an operational interpretation as the minimal cost of the local reconstruction of the quantum observable $A$ through the measurement $M$ over the state $\rho$.  To expound on this, let the act of local reconstruction be implemented by the pullback $\dpb{M}{\rho}$ of the measurement, which creates localised quantum observables $\dpb{M}{\rho} f \in \qob{H}{\rho}$ out of localised classical observables (or, estimators) $f \in \cob{\Omega}{M\!\rho}$.  The precision of the local reconstruction shall be evaluated by the gauge
\begin{equation}\label{def:gauge}
\varepsilon_{\rho}(A,f;M)
	\defeq \sqrt{ \lnorm{ A - \dpb{M}{\rho}f }{\rho}^{2} + \Bigl( \lnorm{ f }{M\!\rho}^{2} - \lnorm{ \dpb{M}{\rho}f }{\rho}^{2} \Bigr) }
\end{equation}
in this paper.  Here, the first term $\lnorm{ A - \dpb{M}{\rho}f }{\rho}$ gives an evaluation of the algebraic deviation between the target observable $A$ to be reconstructed and the operator $\dpb{M}{\rho}f$ created by means of the measurement $M$ out of $f$, whereas the second term $\lnorm{ f }{M\!\rho}^{2} - \lnorm{ \dpb{M}{\rho}f }{\rho}^{2} = \lstdv{f}{M\!\rho}^{2} - \lstdv{\dpb{M}{\rho}f}{M\!\rho}^{2} \geq 0$ represents the potential increase of the cost in the reconstruction itself owing to the suboptimal choice of either or both of the measurement $M$ and the function $f$ (see discussions below the equality \eqref{ineq:non-expansiveness_adjoint}).

A useful observation is that the square of the gauge \eqref{def:gauge} admits the decomposition
\begin{equation}\label{eq:decomposition_gauge}
\varepsilon_{\rho}(A,f;M)^{2}
	= \err{A}{M}{\rho}^{2} + \lnorm{ \dpf{M}{\rho}A - f }{M\!\rho}^{2}
\end{equation}
into the sum of the squares of the error \eqref{def:error_quantum-measurement} and the potential cost induced by the suboptimality of the choice of $f$;  this could be confirmed by simple computation using \eqref{char:pullback_pushforward}.

At this point, one immediately finds that the classical observable that minimises the equality \eqref{eq:decomposition_gauge} is furnished by the pushforward $f = \dpf{M}{\rho}A$;  this offers an operational characterisation of the error
\begin{equation}\label{char:error}
\min_{f} \varepsilon_{\rho}(A,f;M)
	= \err{A}{M}{\rho}
\end{equation}
as the minimum of the gauge over all $f$ in reconstructing the quantum observable $A$, along with the interpretation of the pushforward as the unique (up to equivalence) optimal function that attains it.

It is also worth noting that, in view of the decomposition $\err{A}{M}{\rho}^{2} = \lstdv{A}{\rho}^{2} - \lstdv{\pf{M}{\rho}A}{M\!\rho}^{2}$, the error is always bounded from above by the standard deviation
\begin{equation}
\lstdv{A}{\rho} \geq \err{A}{M}{\rho} \geq 0,
\end{equation}
and is therefore bounded with respect to the seminorm $\lnorm{A}{\rho}$ on $\qob{H}{\rho}$.

\subsection{Error of a Quantum Measurement under Local Representability}

In this paper, an observable $A \in \ran{\pb{M}{\rho}}$ is said to be \emph{locally representable} by a quantum measurement $M$ over the state $\rho$ (or briefly, $\rho$-representable), if it belongs to the image of the pullback.  Under such a situation, a classical observable $f$ satisfying $\pb{M}{\rho}f = A$ shall be called a \emph{local representative} of $A$ regarding $M$ over $\rho$ (or briefly, a $\rho$-representative).  Intuitively, locally representable observables may be interpreted as those admitting local reconstructions by classical observables regarding the said measurement.

\subsubsection{Error for Local Representability}

The condition of local representability leads to another natural definition of the error.  The author defines the \emph{error for local representability} regarding a measurement $M$ of an observable $A \in \ran{\pb{M}{\rho}}$ that is locally representable over $\rho$ by the amount of expansion\begin{equation}\label{def:error_quantum-measurement_representability}
\rerr{A}{M}{\rho}
	\defeq \sqrt{\lnorm{\invpb{M}{\rho}A}{M\!\rho}^{2} - \lnorm{A}{\rho}^{2}}
\end{equation}
induced by the partial inverse of the pullback.  Given the boundedness---and hence the closedness---of the pullback, its partial inverse is a well-defined closed operator, and consequently the error \eqref{def:error_quantum-measurement_representability} is well-defined.

Note that, while the pullback is always bounded, its partial inverse is in general not necessarily so; as such, the error \eqref{def:error_quantum-measurement_representability} may be unbounded.  The characterisation of the condition on which the partial inverse of a closed operator is bounded is dictated by the renowned closed graph theorem;  one finds form its straightforward application that the partial inverse of a closed operator is bounded if and only if its domain is topologically closed (which, in this context, translates to the condition $\ran{\pb{M}{\rho}} = \cran{\pb{M}{\rho}}$), or equivalently, if the closed operator has a closed range.  It goes without saying that the same properties apply to the partial inverse of the pushforward as well.

As with the error \eqref{def:error_quantum-measurement} introduced earlier, non-negativity $\rerr{A}{M}{\rho} \geq 0$ of the error \eqref{def:error_quantum-measurement_representability} for local representability follows from the non-expansiveness of the pushforward combined with the property \eqref{char:partial-inverse_02} of the partial inverse.  It is also straightforward to check the absolute homogeneity $\rerr{tA}{M}{\rho} = \abs{t}\, \rerr{A}{M}{\rho}$, $\forall t \in \mathbb{R}$, and the subadditivity $\rerr{A}{M}{\rho} + \rerr{B}{M}{\rho} \geq \rerr{A+B}{M}{\rho}$.  In other words, the error \eqref{def:error_quantum-measurement_representability} also furnishes a seminorm on the subspace $\ran{\pb{M}{\rho}} \subset \qob{H}{\rho}$ of all the $\rho$-representable observables by the measurement $M$.

\subsubsection{Closedness of the Error}

Closedness of the partial inverse entails closedness of the error \eqref{def:error_quantum-measurement_representability} as (the square root of) a quadratic form, which is another noteworthy and convenient property of the novel construction of error presented in this paper.

Recall that a symmetric (Hermitian) sesquilinear form (or briefly, a symmetric form) $\mathfrak{t}$ defined on a subspace $D_{\mathfrak{t}}$ of a Hilbert space $\HS{H}$ is called lower semibounded, if there exists a real number $m \in \mathbb{R}$ such that $\mathfrak{t}[x,x] \geq m \norm{x}^{2}$ holds for all $x \in D_{\mathfrak{t}}$, where $\norm{\,\cdot\,}$ is the norm of the original Hilbert space $\HS{H}$.  In such a case, the number $m$ is called a lower bound for $\mathfrak{t}$, and it is customary to write $\mathfrak{t} \geq m$.  If the form has $0$ as a lower bound, \textit{i.e.}, $\mathfrak{t} \geq 0$, it is called positive.  A lower semibounded symmetric form $\mathfrak{t}$ defined on a dense subspace $D_{\mathfrak{t}}$ of a Hilbert space is called closed, if the space $D_{\mathfrak{t}}$ is complete with respect to the inner product $\linpr{x}{y}{\mathfrak{t}} \defeq \mathfrak{t}[x,y] + (1 - m) \inpr{x}{y}$, where $\inpr{\,\cdot\,}{\,\cdot\,}$ is the inner product of the original Hilbert space $\HS{H}$.

Given the fact that the error \eqref{def:error_quantum-measurement_representability} for local representability fulfils the parallelogram identity, which could be readily confirmed, it induces a unique positive symmetric form associated with it.  With rudimentary techniques of functional analysis, it is then straightforward to find by construction that the closedness of the partial inverse entails the closedness of the positive symmetric form uniquely associated with the error \eqref{def:error_quantum-measurement_representability}.

\subsubsection{Lower Semicontinuity of the Error}

A useful characterisation of closedness of forms is known in terms of lower semicontinuity.  Recall that a function $f : X \to \mathbb{R} \cup \{\infty\}$ on a metric space $X$ is called lower semicontinuous, if $\liminf_{n\to\infty} f(x_{n}) \geq f(\lim_{n\to\infty} x_{n})$ holds for any convergent sequence $x_{n}$ in $X$.  Under the same settings as above, with a slight abuse of notation, let $\mathfrak{t}[x] \defeq \mathfrak{t}[x,x]$ denote the quadratic form associated with a densely defined lower semibounded symmetric form $\mathfrak{t}$, and consider its extension $\mathfrak{t}^{\prime} : \HS{H} \to \mathbb{R} \cup \{\infty\}$ to the whole Hilbert space by letting $\mathfrak{t}^{\prime}[x] \defeq \mathfrak{t}[x]$ for $x \in D_{\mathfrak{t}}$ and $\mathfrak{t}^{\prime}[x] \defeq +\infty$ otherwise.  Then, it is well-known that $\mathfrak{t}$ is closed if and only if its extension $\mathfrak{t}^{\prime}$ is lower semicontinuous.

In the current context of quantum measurements, one may thus extend the original definition \eqref{def:error_quantum-measurement_representability} of error for local representability to the completion (or equivalently, topological closure) of the original domain by defining
\begin{equation}\label{def:extension_error_quantum-measurement_representability}
\rerr{A}{M}{\rho}
	\defeq
		\begin{dcases}
		\rerr{A}{M}{\rho} & (A \in \ran{\pb{M}{\rho}}), \\
		+\infty & (A \notin \ran{\pb{M}{\rho}}),
		\end{dcases}
\end{equation}
where a slight abuse of notation is made by denoting the original definition and its extension with the same symbol.

Given the closedness of the error \eqref{def:error_quantum-measurement_representability} as (the square root of) a quadratic form, it is then immediate from the aforementioned equivalence that the extended error \eqref{def:extension_error_quantum-measurement_representability} is lower semicontinuous, \textit{i.e.}, for any convergent $A_{n} \to A$ sequence of localised quantum observables in the extended domain, the inequality
\begin{equation}
\liminf_{n\to\infty} \rerr{A_{n}}{M}{\rho}
	\geq \rerr{A}{M}{\rho}
\end{equation}
holds.

While the following passages of this paper focus mostly on the original definition \eqref{def:error_quantum-measurement_representability} of the error for the sake of simplicity of arguments, many of the results and properties presented there remain valid for the extended definition \eqref{def:extension_error_quantum-measurement_representability} as well with the usual mathematical conventions regarding the infinity.  Further extension of the definition \eqref{def:error_quantum-measurement_representability} to the whole Hilbert space $\qob{H}{\rho}$ is also possible;  it will be addressed elsewhere in an appropriate context.

\subsubsection{Operational Interpretation}

As with the definition \eqref{def:error_quantum-measurement}, the error \eqref{def:error_quantum-measurement_representability} for local representability also admits an operational interpretation as the minimal cost of the local reconstruction of the quantum observable $A \in \ran{\pb{M}{\rho}}$, albeit under a different constraint.

A useful observation for the argument is that, under the constraint $\pb{M}{\rho}f = A$ regarding local representability, the square of the gauge \eqref{def:gauge} admits a decomposition
\begin{equation}\label{eq:decomposition_gauge_representability}
\varepsilon_{\rho}(A,f;M)^{2} = \rerr{A}{M}{\rho}^{2} + \lnorm{ \invpb{M}{\rho}A - f }{M\!\rho}^{2}
\end{equation}
into the sum of the squares of the error \eqref{def:error_quantum-measurement_representability} and the potential cost of the suboptimal choice of the local representative;  this may be readily confirmed by simple computation utilising the construction and the minimality property \eqref{char:minimality_argmin} of the partial inverse.

At this point, it is easy to see that the $\rho$-representative that minimises the gauge \eqref{eq:decomposition_gauge_representability} is furnished by the partial inverse $f = \invpb{M}{\rho} A$ regarding the quantum observable of interest;  this offers an operational characterisation of the error
\begin{equation}\label{char:error_representability}
\min_{\pb{M}{\rho}f = A} \varepsilon_{\rho}(A,f;M)
	= \rerr{A}{M}{\rho}
\end{equation}
as the minimum of the gauge over all the classical observables $f$ in representing the quantum observable $A$ under the constraint $\pb{M}{\rho}f = A$ of local representability, along with the interpretation of the partial inverse as the unique (up to equivalence) locally optimal function that attains it.

For another explicit interpretation, given the fact that the expectation value is invariant under the pullback and the pushforward as mentioned earlier (it shall be also noted here that, as a corollary to this property, the partial inverse $\lev{ \invpb{M}{\rho} A }{M\!\rho} = \lev{A}{\rho}$ of the pullback and that $\lev{ \invpf{M}{\rho} f }{\rho} = \lev{f}{M\!\rho}$ of the pushforward also preserve the expectation value), observe that the equality \eqref{eq:decomposition_gauge_representability} may be rewritten into
\begin{equation}\label{eq:decomposition_standard-deviation_representability}
\lstdv{f}{M\!\rho}^{2}
	=  \lstdv{A}{\rho}^{2} + \rerr{A}{M}{\rho}^{2} + \lnorm{ \invpb{M}{\rho}A - f }{M\!\rho}^{2}.
\end{equation}
This admits an interpretation as a decomposition of the variance of a $\rho$-representative $f$ of an observable $A$ into the sums of the squares of the quantum fluctuation (quantum standard deviation) of $A$, that of the error \eqref{def:error_quantum-measurement_representability}, and that of the contribution of the suboptimal choice of the function $f$.

Specifically, note that the first component $\lstdv{A}{\rho}$ of the decomposition \eqref{eq:decomposition_standard-deviation_representability} is determined solely by the choice of the quantum observable and the state, thereby setting a universal lower bound to the statistical cost $\lstdv{f}{M\!\rho}$ of the reconstruction of the quantum observable $A$ of interest, whereas the second component $\rerr{A}{M}{\rho}$ offers an additional contribution to the fundamental bound imposed by the choice of the measurement $M$.  In this regard, it is beneficial to note a trivial corollary
\begin{equation}\label{ev:standard-deviation_lower_bound}
\lstdv{f}{M\!\rho}^{2} \geq \lstdv{A}{\rho}^{2} + \rerr{A}{M}{\rho}^{2} = \dlstdv{\invpb{M}{\rho} A}{M\!\rho}^{2}
\end{equation}
to the decomposition \eqref{eq:decomposition_standard-deviation_representability} in the form of an inequality, in which the role of the two contributors to the lower bound becomes apparent.

The third component $\lnorm{ \invpb{M}{\rho}A - f }{M\!\rho}$ of the decomposition \eqref{eq:decomposition_standard-deviation_representability} furthermore takes into account the additional cost induced by the potentially suboptimal choice of the local representative $f$.  As is familiar, the choice of the function $f$ effectively corresponds to the choice of the method of processing the measurement outcomes, and as such is often the primary location in which one seeks for optimisation in many real-world situations.  The decomposition \eqref{eq:decomposition_standard-deviation_representability} dictates that the optimal choice of the representative is uniquely given by the partial inverse $f = \invpb{M}{\rho} A$, in which case the lower bound \eqref{ev:standard-deviation_lower_bound} of the statistical cost is uniquely attained;  this fact may alternatively be directly obtained in view of the minimality \eqref{char:minimality} of the partial inverse $\min_{\pb{M}{\rho}f = A} \lnorm{f}{M\!\rho} = \lnorm{\invpb{M}{\rho}A}{M\!\rho}$, which is equivalent to
\begin{equation}
\min_{\pb{M}{\rho}f = A} \lstdv{f}{M\!\rho} = \lstdv{\invpb{M}{\rho}A}{M\!\rho}
\end{equation}
given the fact that the pullback and the pushforward (as well as their partial inverses) preserve the expectation value.

\subsection{Two Definitions of Error}

The relation between the two definitions \eqref{def:error_quantum-measurement} and \eqref{def:error_quantum-measurement_representability} of the error may be explicitly addressed by the equality
\begin{equation}\label{eq:two_errors}
\rerr{A}{M}{\rho}^{2} - \err{A}{M}{\rho}^{2}
	= \lnorm{\invpb{M}{\rho}A - \dpf{M}{\rho}A}{\rho}^{2}.
\end{equation}
One direct way to confirm this is through a simple application of the characterisation \eqref{char:pullback_pushforward} regarding the adjoints of the pullback and pushforward, and that \eqref{eq:partial-inverse_adjoint} of their partial inverses, the latter of which explicitly reads
\begin{equation}
\invpf{M}{\rho} = \bigl(\invpb{M}{\rho}\bigr)^{*}
\end{equation}
under the current context;  another trivial method would be to either substitute $f = \invpb{M}{\rho} A$ in the equality \eqref{eq:decomposition_gauge}, or to substitute $f = \pf{M}{\rho} A$ in the equality \eqref{eq:decomposition_gauge_representability}.   Note that the above relation \eqref{eq:two_errors} specifically entails the evaluation
\begin{equation}\label{eval:two_errors}
\rerr{A}{M}{\rho} \geq \err{A}{M}{\rho}
\end{equation}
given the positivity of the seminorm on the right-hand side; another simple way to obtain this evaluation is to compare the characterisations \eqref{char:error} and \eqref{char:error_representability} of the two errors regarding the two different strategies---the latter involves the constraint $\pb{M}{\rho}f = A$ regarding local representability, whereas the former is free from such constraint---in optimising the gauge \eqref{def:gauge}.  In fact, one finds in the next passage that the equality of the two definition holds precisely when one of the definitions of the error vanishes, in which case they both necessarily vanish together.

\subsection{Errorless Measurement\label{sec:errorless_measurement}}

The identification of the conditions on which the measurement becomes free from error is of great importance.  In this paper, a quantum measurement $M$ is said to be capable of an errorless measurement of $A$ over $\rho$ (under local representability), if (the observable $A \in \ran{\pb{M}{\rho}}$ is locally representable and) the error \eqref{def:error_quantum-measurement} (the error \eqref{def:error_quantum-measurement_representability} for local representability) vanishes.  Among the several characterisations of the errorless measurement, which shall be expounded in detail elsewhere, the following five equivalent conditions that are most relevant to the context of this paper shall be presented below:
\begin{enumerate}[label=\rm{(\alph*)}]
\item $\rerr{A}{M}{\rho} = 0$,\label{def:errorless-measurement_representability}
\item $A = \invpf{M}{\rho}\invpb{M}{\rho}A$,\label{char:errorless-measurement_1}
\item $\rerr{A}{M}{\rho} = \err{A}{M}{\rho}$,\label{char:errorless-measurement_2}
\item $A = \dpb{M}{\rho}\dpf{M}{\rho}A$,\label{char:errorless-measurement_3}
\item $\err{A}{M}{\rho} = 0$.\label{def:errorless-measurement}
\end{enumerate}
Here, a harmless abuse of expression is made in order to avoid clutter, in which obvious preconditions are implicitly assumed;  it goes without saying that, in precise terms, the condition \ref{def:errorless-measurement_representability} should be understood as `$A \in \ran{\pb{M}{\rho}}$ and $\rerr{A}{M}{\rho} = 0$', condition \ref{char:errorless-measurement_1} as `$A \in \ran{\pb{M}{\rho}}$, $\invpb{M}{\rho}A \in \ran{\pf{M}{\rho}}$, and $A = \invpf{M}{\rho}\invpb{M}{\rho}A$', and condition \ref{char:errorless-measurement_2} as `$A \in \ran{\pb{M}{\rho}}$ and $\rerr{A}{M}{\rho} = \err{A}{M}{\rho}$' in order to guarantee their well-definedness in the first place.

A simple proof of the above equivalences reads as follows: $\ref{def:errorless-measurement_representability} \implies \ref{def:errorless-measurement}$ is trivial from the evaluation \eqref{eval:two_errors}, $\ref{def:errorless-measurement} \implies \ref{char:errorless-measurement_3}$ is an immediate consequence of the characterisation \eqref{eq:decomposition_gauge}, from which $\err{A}{M}{\rho} \geq \lnorm{ A - \dpb{M}{\rho}\dpf{M}{\rho} A }{\rho}$ follows, $\ref{char:errorless-measurement_3} \implies \ref{char:errorless-measurement_1}$ may be readily confirmed by applying $\invpf{M}{\rho}\invpb{M}{\rho}$ on both-hand sides of the assumption, $\ref{char:errorless-measurement_1} \implies \ref{char:errorless-measurement_2}$ is an immediate consequence of \eqref{eq:two_errors} with a simple substitution of the assumption, and $\ref{char:errorless-measurement_2} \implies \ref{def:errorless-measurement_representability}$ is immediate from \eqref{eq:two_errors}, which yields $\invpb{M}{\rho}A = \dpf{M}{\rho}A$, and thus results in $\lnorm{A}{\rho} \geq \lnorm{\dpf{M}{\rho}A}{M\!\rho} = \lnorm{\invpb{M}{\rho}A}{M\!\rho} \geq \lnorm{A}{\rho}$ given the consequences of the non-expansiveness of the pullback and the pushforward.

Specifically, the above equivalences reveal that the conditions on which the two definitions \eqref{def:error_quantum-measurement} and \eqref{def:error_quantum-measurement_representability} of the error vanish is identical.  In this regard, while the former definition is numerically always less than the latter for local representability barring the errorless case (see discussions around the equality \eqref{eval:two_errors}), the two definitions of the error introduced by the author can be understood as being equivalent for the purpose of detecting the condition on which the measurement becomes free from error.

Another consequence worthy of mention is that an errorless measurement of a self-adjoint observable is always available; the projection measurement associated with it is capable of such a measurement \emph{globally}, \textit{i.e.}, over every state $\rho$.  A simple proof would be to observe that the pushforward of an observable $A$ by the projection measurement $M$ associated with reads $\pf{M}{\rho}A = \mathrm{id}$, where $\mathrm{id}$ is the identity function on the spectrum of $A$;  this may be directly computed based on the formula \eqref{eq:pushforward_Radon-Nikodym}, but may also---perhaps more effortlessly---be confirmed by verifying that the identity function indeed fulfils the characterisation \eqref{char:pullback_pushforward} of the pushforward.  This fact, combined with the condition \ref{char:errorless-measurement_3} for example, shall lead to the desired statement.

\section{Uncertainty Relations for Errors\label{sec:uncertainty_relation}}

Now that the necessary tools and concepts have been introduced, the new uncertainty relations, which are the main results of this paper, shall be presented.  In fact, several different inequalities are available that mark the trade-off relation regarding each definition \eqref{def:error_quantum-measurement} and \eqref{def:error_quantum-measurement_representability} of the errors, but the simplest among them, which should suffice for the purpose of this paper,  shall be presented below.

\subsection{Uncertainty Relation for Errors}

Let $A$ and $B$ be an arbitrary pair of quantum observables of the system $\mathcal{H}$, and also let $\rho \in Z(\mathcal{H})$ be a quantum state of one's choice.  Then, for any quantum measurement $M: Z(\mathcal{H}) \to W(\Omega)$, the inequality
\begin{equation}\label{ineq:urel_error}
\err{A}{M}{\rho}\, \err{B}{M}{\rho}
	\geq \sqrt{\mathcal{R}^2 + \mathcal{I}^2}
\end{equation}
holds with 
\begin{equation}\label{def:urel_real}
\mathcal{R}
	\defeq \dlev{ \frac{\{A,B\}}{2} }{\!\!\rho} - \dlinpr{ \dpf{M}{\rho}A }{ \dpf{M}{\rho}B }{M\!\rho}
\end{equation}
and
\begin{equation}\label{def:urel_imaginary}
\mathcal{I}
	\defeq \dlev{ \frac{[A,B]}{2i} }{\!\!\rho}  - \dlev{ \frac{[\dpb{M}{\rho}\dpf{M}{\rho}A,B]}{2i} }{\!\!\rho} - \dlev{ \frac{[A,\dpb{M}{\rho}\dpf{M}{\rho}B]}{2i} }{\!\!\rho}
\end{equation}
being the two contributors to the lower bound.

The proof of the inequality \eqref{ineq:urel_error} is actually quite simple; it is just a direct corollary to the renowned Cauchy--Schwarz inequality.  In fact, with the help of the semi-inner product (\textit{i.e.}, positive semi-definite symmetric sesquilinear form)
\begin{equation}\label{def:aux_semi-inner_product}
\inpr{ (X, f) }{ (Y, g) }
	\defeq \lev{ X^{\dagger}Y }{\rho} + \lev{ f^{\dagger} g }{M\!\rho} - \lev{ M^{\prime}f^{\dagger} M^{\prime}g }{\rho}
\end{equation}
defined for the products of Hilbert-space operators $X$ and complex functions $f$, one finds that the error $\err{A}{M}{\rho} = ( \inpr{ (X_{A}, f_{A}) }{ (X_{A}, f_{A}) } )^{1/2}$ admits a description by the induced seminorm with the shorthands $X_{A} \defeq A - \dpb{M}{\rho}\dpf{M}{\rho}A$ and $f_{A} \defeq \dpf{M}{\rho}A$.  The Cauchy--Schwarz inequality applied to the seminorm thus reveals that the product of the errors
\begin{equation}
\err{A}{M}{\rho}\, \err{B}{M}{\rho}
	\geq \abs{ \inpr{ (X_{A}, f_{A}) }{ (X_{B}, f_{B}) } }
\end{equation}
is bounded from below by the absolute value of the semi-inner product $\inpr{ (X_{A}, f_{A}) }{ (X_{B}, f_{B}) } = \mathcal{R} + i\,\mathcal{I}$, whose real part $\mathcal{R}$ is given by \eqref{def:urel_real} whereas the imaginary part $\mathcal{I}$ by \eqref{def:urel_imaginary}.  This completes the proof of the inequality \eqref{ineq:urel_error}.

From a mathematical (geometric) point of view, the real part $\mathcal{R}$ in \eqref{def:urel_real} represents the decrease of the induced metric on the bundle of `localised' quantum observables, whose loss being inevitably caused by the quantum measurement $M$.  In fact, this term is also found to be shared with the inequality regarding the errors of classical measurements $K: W(\Omega) \to W(\Omega^{\prime})$, which are defined as affine maps between classical-state spaces (see FIG.~\ref{fig:quantum_and_classical_measurements}).  With analogous definitions and parallel arguments, one finds that the product of the errors
\begin{equation}\label{ineq:classical_uncertainty-relation_error}
\err{a}{K}{p}\, \err{b}{K}{p}\,
	\geq \abs{ \mathcal{R} }
\end{equation}
of two classical observables ({\it i.e.}, real functions on the sample space $\Omega$) $a(\omega)$ and $b(\omega)$ by the measurement $K$ over $p \in W(\Omega)$ is bounded from below by the absolute value of
\begin{equation}\label{def:classical_urel_real}
\mathcal{R}
	= \dlinpr{ a }{ b }{p} - \dlinpr{ \pf{K}{p}a }{ \pf{K}{p}b }{K\!p},
\end{equation}
which correspond to the real part \eqref{def:urel_real} regarding quantum measurements.  This fact reveals that the semiclassical contributor $\mathcal{R}$ of the lower bound is shared in common between both classical and quantum measurements, thereby suggesting that it is not necessarily of quantum origin.

On the other hand, the imaginary part $\mathcal{I}$ in \eqref{def:urel_imaginary}, which consists of three commutators and imposes an additional constraint on the attainable lower bound of the product of the errors, marks the essence of quantum measurements.  In view of this, the simplified form
\begin{equation}\label{ineq:uncertainty-relation_error_simplified}
\err{A}{M}{\rho}\, \err{B}{M}{\rho}
	\geq \abs{ \mathcal{I} }
\end{equation}
of the relation, which can be trivially obtained by omitting the semiclassical contributor $\mathcal{R}$ in the relation \eqref{ineq:urel_error}, should be mostly adequate to account for the distinguishing characteristics of quantum theory.

\subsection{Uncertainty Relation for Errors under Local Representability}

Under the same settings as above, assume furthermore that the pair of observables $A, B \in \ran{\pb{M}{\rho}}$ are locally representable with respect to the quantum measurement $M$ over the state $\rho$.  Then, the inequality
\begin{equation}\label{ineq:urel_error_representability}
\rerr{A}{M}{\rho}\, \rerr{B}{M}{\rho}
	\geq \sqrt{\tilde{\mathcal{R}}^2 + \mathcal{I}_{0}^2}
\end{equation}
holds with 
\begin{equation}\label{def:urel_real_representability}
\tilde{\mathcal{R}}
	\defeq \dlev{ \frac{\{A,B\}}{2} }{\!\!\rho} - \dlinpr{ \invpb{M}{\rho}A }{ \invpb{M}{\rho}B }{M\!\rho}
\end{equation}
and
\begin{equation}\label{def:urel_imaginary_representability}
\mathcal{I}_{0}
	\defeq \dlev{ \frac{[A,B]}{2i} }{\!\!\rho}
\end{equation}
being the two contributors to the lower bound.

The proof of the inequality \eqref{ineq:urel_error_representability} is merely a parallel to that of the previous inequality \eqref{ineq:urel_error};  in view of the fact that the error reads $\rerr{A}{M}{\rho} = ( \inpr{ (\tilde{X}_{A}, \tilde{f}_{A}) }{ (\tilde{X}_{A}, \tilde{f}_{A}) } )^{1/2}$ with the shorthands $\tilde{X}_{A} \defeq A - \dpb{M}{\rho}\invpb{M}{\rho}A = 0$ and $\tilde{f}_{A} \defeq \invpb{M}{\rho}A$ under the same notations, the Cauchy--Schwartz inequality for the product of the errors reads
\begin{equation}
\rerr{A}{M}{\rho}\, \rerr{B}{M}{\rho}
	\geq \abs{ \inpr{ (\tilde{X}_{A}, \tilde{f}_{A}) }{ (\tilde{X}_{B}, \tilde{f}_{B}) } }
\end{equation}
with the lower bound now being dictated by the absolute value of $ - \inpr{ (\tilde{X}_{A}, \tilde{f}_{A}) }{ (\tilde{X}_{B}, \tilde{f}_{B}) } = \tilde{\mathcal{R}} + i\,\mathcal{I}_{0}$ (here, the negative sign merely serves an aesthetic purpose), whose real $\tilde{\mathcal{R}}$ and the imaginary $\mathcal{I}_{0}$ parts are respectively given by \eqref{def:urel_real_representability} and \eqref{def:urel_imaginary_representability}.  This completes the proof of the inequality \eqref{ineq:urel_error_representability}.
 
From a geometric point of view, the (negative of the) real part $-\tilde{\mathcal{R}}$ in \eqref{def:urel_real_representability} represents the increase of the induced metric on the bundle of localised quantum observables, whose growth being inevitably caused by the measurement $M$.  As with the previous relation \eqref{ineq:urel_error}, this term is also found to be shared with the inequality regarding errors of classical measurements $K$.  With analogous definitions and parallel arguments, one finds that the product of the errors
\begin{equation}\label{ineq:classical_uncertainty-relation_error_representability}
\rerr{a}{K}{p}\, \rerr{b}{K}{p}\,
	\geq \abs{ \tilde{\mathcal{R}} }
\end{equation}
of two classical observables $a(\omega)$ and $b(\omega)$ by the measurement $K$ over $p \in W(\Omega)$ is bounded from below by the absolute value of
\begin{equation}\label{def:classical_urel_real_representability}
\tilde{\mathcal{R}}
	= \linpr{ a }{ b }{p} - \dlinpr{ \invpb{K}{p}a }{ \invpb{K}{p}b }{K\!p},
\end{equation}
which correspond to the real part \eqref{def:urel_real_representability} regarding quantum measurements.  This fact suggests that the semiclassical bound $\tilde{\mathcal{R}}$ is not necessarily of quantum origin.

On the other hand, the imaginary part $\mathcal{I}_{0}$ in \eqref{def:urel_imaginary_representability}, which is nothing but the familiar commutator term found in the lower bound of the Kennard--Robertson relation, marks the essence of quantum measurements.  In this regard, the reduced simpler form $\rerr{A}{M}{\rho}\, \rerr{B}{M}{\rho} \geq \lvert \mathcal{I}_{0} \rvert$ obtained from \eqref{ineq:urel_error_representability} should be mostly adequate.

\subsection{Uncertainty Relation for Local Representatives}

The uncertainty relation \eqref{ineq:urel_error_representability} under local representability entails another noteworthy relation regarding the statistical cost of reconstruction as a corollary.  Among several variants of such inequalities, some of which should be readily found from simply examining the passages below, the following relation should be worthy of presentation in that it achieves a decent balance among the important criteria such as simplicity of expression, tightness of inequality, ease of proof, and physical insight.

Under the same assumptions as in the uncertainty relation \eqref{ineq:urel_error_representability}, for any pair of $\rho$-representatives $f$ of $A = \pb{M}{\rho}f$ and $g$ of $B = \pb{M}{\rho}g$, the inequality
\begin{equation}\label{ineq:urel_error_representability_cost}
\lstdv{f}{M\!\rho}\, \lstdv{g}{M\!\rho}
	\geq \sqrt{ \left( \abs{\tilde{\mathcal{R}}} + \abs{\mathcal{R}_{0}} \right)^{2} + 4\,\mathcal{I}_{0}^{2} }
\end{equation}
holds.  Here, the contributors $\tilde{\mathcal{R}}$ and $\mathcal{I}_{0}$ to the lower bound are respectively given by \eqref{def:urel_real_representability} and \eqref{def:urel_imaginary_representability}, and the contributor
\begin{equation}\label{def:quantum-covariance}
\mathcal{R}_{0}
	\defeq \dlev{ \frac{\{A,B\}}{2} }{\!\!\rho} - \lev{A}{\rho} \lev{B}{\rho}
\end{equation}
is the familiar quantum covariance of the pair of observables.

A simple proof would be to reiteratively apply the Cauchy--Schwarz inequality to the right-hand side of the inequality \eqref{ev:standard-deviation_lower_bound}, thereby obtaining
\begin{align}
&\lstdv{f}{M\!\rho}^{2}\, \lstdv{g}{M\!\rho}^{2} \notag \\
	&\quad \geq \dabs{ \rerr{A}{M}{\rho}\,\rerr{B}{M}{\rho} + \lstdv{A}{\rho}\,\lstdv{B}{\rho} }^{2} \notag \\
	&\quad \geq \left( \tilde{\mathcal{R}}^{2} + \mathcal{I}_{0}^{2} \right) + 2 \sqrt{ \tilde{\mathcal{R}}^{2} + \mathcal{I}_{0}^{2} } \sqrt{ \mathcal{R}_{0}^{2} + \mathcal{I}_{0}^{2} }  + \left( \mathcal{R}_{0}^{2} + \mathcal{I}_{0}^{2} \right) \notag \\
	&\quad \geq \left( \tilde{\mathcal{R}}^{2} + \mathcal{I}_{0}^{2} \right) + 2 \left(\abs{ \tilde{\mathcal{R}} } \abs{ \mathcal{R}_{0} } + \mathcal{I}_{0}^{2} \right)  + \left( \mathcal{R}_{0}^{2} + \mathcal{I}_{0}^{2} \right) \notag \\
	&\quad = \left( \abs{\tilde{\mathcal{R}}} + \abs{\mathcal{R}_{0}} \right)^{2} + 4\,\mathcal{I}_{0}^{2}.
\end{align}
Here, the second inequality is due to the uncertainty relation \eqref{ineq:urel_error_representability} and the Schr{\"o}dinger relation \cite{Schroedinger_1930}
\begin{equation}\label{ineq:uncertainty-relation_Schroedinger}
\lstdv{A}{\rho}\, \lstdv{B}{\rho} \geq \sqrt{\mathcal{R}_{0}^2 + \mathcal{I}_{0}^2},
\end{equation}
which are both in essence corollaries to the Cauchy--Schwarz inequality;  in fact, the Schr{\"o}dinger relation is itself found to be a corollary to the formulation of this paper for the trivial choice of the measurement (see Sec.~\ref{sec:reference_to_other_relations}).

It goes without saying that the minimum of the left-hand side of the uncertainty relation \eqref{ineq:urel_error_representability_cost} is attained by the optimal choice of the $\rho$-representative, \textit{i.e.}, the partial inverse, which results in
\begin{equation}
\lstdv{\invpb{M}{\rho}A}{M\!\rho}\, \lstdv{\invpb{M}{\rho}B}{M\!\rho}
	\geq \sqrt{ \left( \abs{\tilde{\mathcal{R}}} + \abs{\mathcal{R}_{0}} \right)^{2} + 4\,\mathcal{I}_{0}^{2} }
\end{equation}
as a trivial corollary.

With parallel definitions and arguments, the classical analogue of the inequality \eqref{ineq:urel_error_representability_cost} may be shown to read
\begin{equation}
\lstdv{f}{K\!p}\, \lstdv{g}{K\!p}
	\geq \abs{\tilde{\mathcal{R}}} + \abs{\mathcal{R}_{0}}
\end{equation}
regarding the $\rho$-representatives $f$ of $a(\omega) = (\pb{K}{p}f)(\omega)$ and $g$ of $b(\omega) = (\pb{K}{p}g)(\omega)$ by the measurement $K$ over $p \in W(\Omega)$.  Here, the contributors to the lower bound are given by 
\begin{equation}
\tilde{\mathcal{R}}
	= \linpr{ a }{ b }{p} - \dlinpr{ \invpb{K}{p}a }{ \invpb{K}{p}b }{K\!p}
\end{equation}
and
\begin{equation}
\mathcal{R}_{0}
	= \linpr{ a }{ b }{p} - \lev{a}{p} \lev{b}{p},
\end{equation}
which respectively correspond to \eqref{def:urel_real_representability} and \eqref{def:quantum-covariance} regarding quantum measurements.

\section{Local Joint-Describability\label{sec:local_joint-describability}}

The author also presents refinements of the uncertainty relations \eqref{ineq:urel_error}, \eqref{ineq:urel_error_representability}, and \eqref{ineq:urel_error_representability_cost} for the cases in which the quantum measurements may be chosen differently for the pair of observables to be measured.

\subsection{Local Joint Description of Quantum Measurements}

In this paper, a set of quantum measurements $M_{i} : Z(\HS{H}) \to W(\Omega_{i})$, $i = 1, 2, \dots, N$ ($N \in \mathbb{N}$), is said to admit a \emph{local joint description} over the state $\rho \in Z(\HS{H})$ (or briefly, $\rho$-joint description), if there exists a mediating quantum measurement $J : Z(\HS{H}) \to W(\Omega)$ and a set of classical processes (measurements) $K_{i} :  W(\Omega) \to  W(\Omega_{i})$ for which either of the two equivalent conditions
\begin{equation}\label{def:joint-description_pb}
\pb{(M_{i})}{\rho} = \pb{J}{\rho} \circ \pb{(K_{i})}{(J\!\rho)},
\end{equation}
or
\begin{equation}\label{def:joint-description_pf}
\pf{(M_{i})}{\rho} = \pf{(K_{i})}{(J\!\rho)} \circ \pf{J}{\rho}
\end{equation}
holds for all $i = 1, 2, \dots, N$; here, note that the above conditions implicitly connote $M_{i}\rho = (K_{i} \circ J)\rho$, which is to say that the probability distributions of the measurement outcomes of $M_{i}$ coincide with that of the composite measurements $K_{i} \circ J$ over $\rho$.

\subsection{Local Joint Measurability}

Without essential loss of generality, in order to avoid too much complication, the rest of this paper focuses on a special case of local joint describability in which a pair of measurements $M_{i} : Z(\mathcal{H}) \to W(\Omega_{i})$, $i = 1, 2$, admits a $\rho$-joint description by a mediating measurement $J : Z(\mathcal{H}) \to W(\Omega_{1} \times \Omega_{2})$ with the outcome space $\Omega = \Omega_{1} \times \Omega_{2}$ being the product space of those of $M_{i}$, and the classical processes $K_{i} = \pi_{i}$ given by the projections
\begin{align}
(\pi_{1} p)(\omega_{1})
	&\defeq \int_{\Omega_{2}} p(\omega_{1},\omega_{2})\, d\omega_{2} \label{def:projection_1}, \\
(\pi_{2} p)(\omega_{2})
	&\defeq \int_{\Omega_{1}} p(\omega_{1},\omega_{2})\, d\omega_{1} \label{def:projection_2}
\end{align}
to the marginals.   In this paper, such a pair of measurements $M_{1}$ and $M_{2}$ shall be said to admit a \emph{local joint-measurement} $J$ over $\rho$ (or briefly, $\rho$-joint measurement), which is of course a special case of a local joint description.

\begin{figure}
\centering
\includegraphics[hiresbb,clip,width=0.40\textwidth,keepaspectratio,pagebox=artbox]{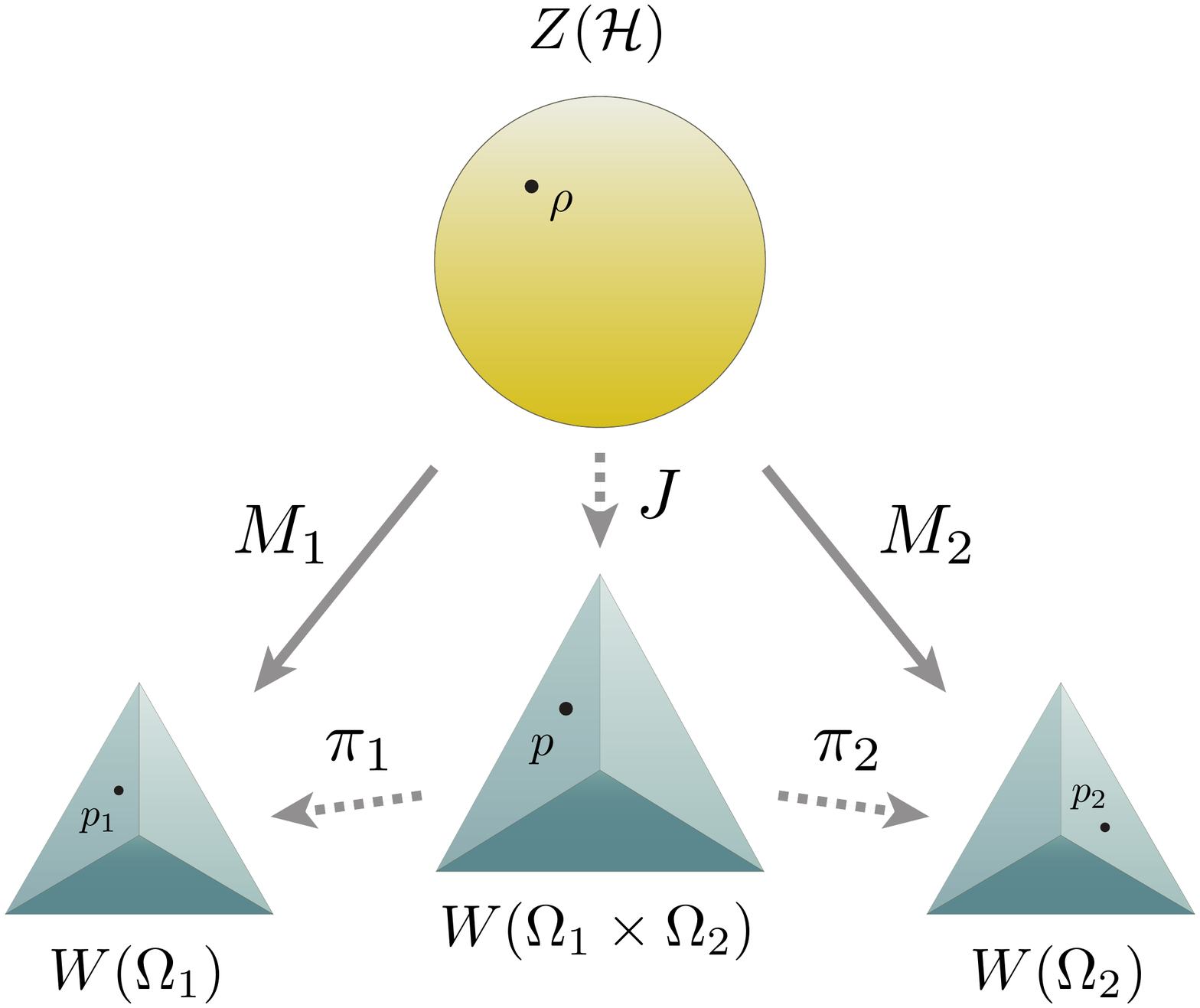}%
\caption{
The basic premise of joint measurability.  A pair of measurements $M_{i} : Z(\mathcal{H}) \to W(\Omega_{i})$, $i=1,2$, are said to admit a (global) joint measurement by $J : Z(\mathcal{H}) \to W(\Omega_{1} \times \Omega_{2})$, if the distributions $p_{i} = M_{i}\rho$ of the outcomes of the measurements $M_{i}$ are given by the respective marginals $p_{i} = \pi_{i}p$ of the joint distribution $p = J\!\rho$ for every state $\rho \in Z(\HS{H})$.
}
\label{fig:joint-measurement}
\end{figure}

An illustrative stronger example of this is when a pair of measurements $M_{i} : Z(\mathcal{H}) \to W(\Omega_{i})$, $i = 1, 2$, admit a \emph{(global) joint measurement} by $J : Z(\mathcal{H}) \to W(\Omega_{1} \times \Omega_{2})$ in the sense that both the distributions $M_{i}\rho$ are given by the marginals of the joint distribution $J\!\rho$ of the joint measurement for all $\rho \in Z(\HS{H})$ (see FIG.~\ref{fig:joint-measurement}).  More explicitly, the maps concerned fulfil the conditions $M_{i} = \pi_{i} \circ J$, thereby satisfying \eqref{def:joint-description_pb}, or equivalently \eqref{def:joint-description_pf}, with the specific classical processes $K_{i} = \pi_{i}$ in their respective places.  

Under the settings of local joint measurability, given the fact that the adjoints of the projections \eqref{def:projection_1} and \eqref{def:projection_2} respectively read $(\pi_{1}^{\prime}f)(x,y) = f(x)$ and $(\pi_{2}^{\prime}g)(x,y) = g(y)$, their pullbacks are found to be isometries, \textit{i.e.}, $\lnorm{ f }{\pi_{1}p} = \lnorm{ \spb{\pi_{1}}f }{p}$ and $\lnorm{ g }{\pi_{2}p} = \lnorm{ \spb{\pi_{2}}g }{p}$.  This allows for the identification of the spaces $R_{M\!\rho}(\Omega_{i})$ regarding each of the measurements with their images under the pullbacks, which are subspaces of the larger space $R_{J\!\rho}(\Omega_{1}\times\Omega_{2})$ of the joint measurement.  For the general case of local describability (in which the pullbacks of the classical processes are not necessarily isometries), the refined uncertainty relations presented below still hold in similar forms with due modifications to the semi-classical contributors \eqref{def:urel_joint_real} and \eqref{def:urel_representability_joint_real} to their lower bounds;  the general case shall be explicated in detail in subsequent papers of the author.

\subsection{Uncertainty Relations under Local Joint-Measurability}

In what follows, only the essence regarding the refined uncertainty relations shall be spelled out in order to avoid unnecessary repetitions.  Various properties of (as well as remarks to) the simple cases \eqref{ineq:urel_error}, \eqref{ineq:urel_error_representability}, and \eqref{ineq:urel_error_representability_cost} are still valid for the respective refined relations \eqref{ineq:urel_error_joint}, \eqref{ineq:urel_error_representability_joint}, and \eqref{ineq:uncert_joint_representability_2} in obvious parallel manners.

Based on the terminologies introduced above, let $A$ and $B$ be quantum observables on $\HS{H}$, and $M : Z(\HS{H}) \to W(\Omega_{1})$ and $N : Z(\HS{H}) \to W(\Omega_{2})$ be quantum measurements.  If the pair of measurements $M$ and $N$ admit a local joint measurement by a quantum measurement $J$ over $\rho \in Z(\HS{H})$, the inequality
\begin{equation}\label{ineq:urel_error_joint}
\err{A}{M}{\rho}\, \err{B}{N}{\rho}
	\geq \sqrt{ \mathcal{R}^{2} + \mathcal{I}^{2} }
\end{equation}
holds.  Here, the contributors to the lower bound are given by the refined terms
\begin{align}\label{def:urel_joint_real}
\mathcal{R}
	&\defeq \dlev{ \frac{\{A,B\}}{2} }{\!\rho} - \dlinpr{ \pf{M}{\rho} A }{ \pf{M}{\rho} B }{M\!\rho} \notag \\
	&\qquad - \dlinpr{ \pf{N}{\rho} A }{ \pf{N}{\rho} B }{N\!\rho} + \dlinpr{ \pf{M}{\rho} A }{ \pf{N}{\rho} B }{J\!\rho}
\end{align}
and
\begin{align}\label{def:urel_joint_imaginary}
\mathcal{I}
	\defeq \dlev{ \frac{[A,B]}{2i} }{\!\rho} - \dlev{ \frac{[\pb{M}{\rho}\pf{M}{\rho}A,B]}{2i} }{\!\rho} - \dlev{ \frac{[A,\pb{N}{\rho}\pf{N}{\rho}B]}{2i} }{\!\rho},
\end{align}
where the abbreviated notation
\begin{equation}
\dlinpr{ \pf{M}{\rho} A }{ \pf{N}{\rho} B }{J\!\rho}
	\defeq \dlinpr{ \spb{\pi_{1}} \pf{M}{\rho} A }{ \spb{\pi_{2}} \pf{N}{\rho} B }{J\!\rho}
\end{equation}
is introduced on the ground of the identifications of the classical observables $f \simeq \spb{\pi_{1}}f$ and $g \simeq \spb{\pi_{2}}g$ mentioned earlier.

Under the same settings as in the refined relation \eqref{ineq:urel_error_joint}, assume moreover that the observables $A \in \ran{\pb{M}{\rho}}$ and $B \in \ran{\pb{N}{\rho}}$ are respectively $\rho$-representable with respect to the measurements $M$ and $N$.  Then, the inequality
\begin{equation}\label{ineq:urel_error_representability_joint}
\rerr{A}{M}{\rho}\, \rerr{B}{N}{\rho}
	\geq \sqrt{ \tilde{\mathcal{R}}^{2} + \mathcal{I}_{0}^{2} }
\end{equation}
holds with the contributions to the lower bound being
\begin{equation}\label{def:urel_representability_joint_real}
\tilde{\mathcal{R}}
	\defeq \dlev{ \frac{\{A,B\}}{2} }{\!\rho} - \dlinpr{ \invpb{M}{\rho}A }{ \invpb{N}{\rho}B }{J\!\rho},
\end{equation}
where the abbreviated notation
\begin{equation}
\dlinpr{ \invpb{M}{\rho}A }{ \invpb{N}{\rho}B }{J\!\rho}
	\defeq \dlinpr{ \spb{\pi_{1}} \invpb{M}{\rho} A }{ \spb{\pi_{2}} \invpb{N}{\rho} B }{J\!\rho}
\end{equation}
is introduced on the ground of the identifications mentioned earlier, and $\mathcal{I}_{0}$ being the same as in \eqref{def:urel_imaginary_representability}.

The proofs of the refined relations \eqref{ineq:urel_error_joint} and \eqref{ineq:urel_error_representability_joint} have in essence the same structure as those of the respective simple cases \eqref{ineq:urel_error} and \eqref{ineq:urel_error_representability};  the adjustment essentially amounts to replacing $M$ with $J$ and applying the identification $f \simeq \spb{\pi_{1}}f$ and $g \simeq \spb{\pi_{2}}g$.  In this regard, they are also mere corollaries to the Cauchy--Schwarz inequality applied to the semi-inner product \eqref{def:aux_semi-inner_product};  more explicitly, in view of the fact that the gauge \eqref{eq:decomposition_gauge_representability}, which may itself be interpreted as another definition of error, admits a description $\varepsilon_{\rho}(A,f;M) = ( \inpr{ (X_{(A,f)}, f) }{ (X_{(A,f)}, f) } )^{1/2}$ by the induced seminorm with the shorthand $X_{(A,f)} \defeq A - \pb{M}{\rho}f$, the Cauchy--Schwarz inequality is found to read
\begin{equation}\label{ineq:gauge}
\varepsilon_{\rho}(A,f;M)\, \varepsilon_{\rho}(B,g;N)
	\geq \sqrt{ R^{2} + I^{2} }
\end{equation}
with
\begin{align}
R
	\defeq \dlev{ \frac{\{A,B\}}{2} }{\!\rho} - \dlinpr{ \pf{M}{\rho} A }{ g }{M\!\rho} 
	- \dlinpr{ f }{ \pf{N}{\rho} B }{N\!\rho} + \dlinpr{ f }{ g }{J\!\rho}
\end{align}
and
\begin{align}
I
	\defeq \dlev{ \frac{[A,B]}{2i} }{\!\rho} - \dlev{ \frac{[\pb{M}{\rho}f,B]}{2i} }{\!\rho} - \dlev{ \frac{[A,\pb{N}{\rho}g]}{2i} }{\!\rho}
\end{align}
being the contributors to the lower bound, where the abbreviated notation
\begin{equation}
\dlinpr{ f }{ g }{J\!\rho}
	\defeq \dlinpr{ \spb{\pi_{1}} f }{ \spb{\pi_{2}} g }{J\!\rho}
\end{equation}
is introduced on the ground of the identifications mentioned earlier.  The refined relations \eqref{ineq:urel_error_joint} and \eqref{ineq:urel_error_representability_joint} then follow from the above inequality \eqref{ineq:gauge} with the specific choices of the classical observables (it goes without saying that parallel remarks holds for the simple cases \eqref{ineq:urel_error} and \eqref{ineq:urel_error_representability_joint} as well); the former relation is proven with the choices $f = \pf{M}{\rho}A$ and $g = \pf{N}{\rho}B$, whereas the latter relation is proven with the choices $f = \invpb{M}{\rho}A$ and $g = \invpb{N}{\rho}B$.

The refined relation \eqref{ineq:urel_error_representability_joint} under local representability entails another noteworthy consequence: under the same assumptions as in \eqref{ineq:urel_error_representability_joint}, the inequality
\begin{equation}\label{ineq:uncert_joint_representability_2}
\lstdv{f}{M\!\rho}\, \lstdv{g}{M\!\rho}
	\geq \sqrt{ \left( \abs{\tilde{\mathcal{R}}} + \abs{\mathcal{R}_{0}} \right)^{2} + 4\,\mathcal{I}_{0}^{2} }
\end{equation}
holds for any pair of $\rho$-representatives $f$ of $A = \pb{M}{\rho}f$ and $g$ of $B = \pb{N}{\rho}g$.  Here, the contributors $\tilde{\mathcal{R}}$, $\mathcal{R}_{0}$, and $\mathcal{I}_{0}$ to the lower bound are respectively given by \eqref{def:urel_representability_joint_real}, \eqref{def:quantum-covariance}, and \eqref{def:urel_imaginary_representability}.

\section{The Uncertainty Principle\label{sec:uncertainty-principle}}

The uncertainty relation \eqref{ineq:urel_error_joint} (and its simple form \eqref{ineq:urel_error}) connotes a potential violation of the na{\"i}ve lower bound $\abs{ \lev{ [A, B] }{\rho} }/2$ for certain choices of quantum measurements; indeed, an errorless measurement of either observable, which is always attainable by the projection measurement associated with it, trivially violates it.  Another way to see this would be to observe that the relation \eqref{ineq:urel_error_joint} is tighter than Ozawa's relation for joint POVM measurements (as will be discussed shortly);  since the latter relation is known to violate the bound, the former does so whenever the latter does.

Nevertheless, Heisenberg's seminal philosophy of the uncertainty principle still remains valid, albeit perhaps in a weaker sense than was originally intended; the uncertainty relation \eqref{ineq:urel_error_joint} reveals a no-go theorem, which states that, whenever $\lev{ [A, B] }{\rho} \neq 0$ is non-vanishing for a pair of quantum observables $A$ and $B$, their joint errorless measurement is impossible.  To see this, suppose there were such a pair of measurements admitting a joint local description.  Then, the relation \eqref{ineq:urel_error_joint} combined with the equivalence $\ref{def:errorless-measurement} \iff \ref{char:errorless-measurement_3}$ would lead to a contradiction $0 \geq \sqrt{\abs{ 0 }^{2} + \abs{ \lev{ [A, B] }{\rho}/2i }^{2}}$.  A simple corollary to this is that, for non-trivial (\textit{i.e.}, $\mathrm{dim}(\HS{H}) \geq 2$) quantum systems, there exists no measurement that is capable of measuring every observable errorlessly over every state by itself.

Given the equivalence of the errorless condition between the two definitions \eqref{def:error_quantum-measurement} and \eqref{def:error_quantum-measurement_representability} of the error, the uncertainty principle in the (weaker) sense of this paper regarding the former error is valid verbatim for the latter error for local representability as well.  Meanwhile, the uncertainty relation \eqref{ineq:urel_error_representability_joint} under local representability (and its simple case \eqref{ineq:urel_error_representability}) differs from the relation \eqref{ineq:urel_error_joint} (and its simple case \eqref{ineq:urel_error}) in that it always respects the na{\"i}ve lower bound:  as long as the pair of observables $A$ and $B$ are locally representable by a pair of measurements that admit a local joint measurement, neither of the errors may vanish whenever $\lev{ [A, B] }{\rho} \neq 0$ is non-vanishing (as opposed to the other relation in which one of the errors may does so).

\section{Reference to Other Relations\label{sec:reference_to_other_relations}}

Below, some of the notable uncertainty relations are examined in the new light of the relations presented in this paper.

\subsection{Quantum Indeterminacy\label{sec:quantum-indeterminacy}}

The celebrated Kennard--Robertson relation, which has very little to do with the concept of measurement, in fact emerges as a trivial case of the relation \eqref{ineq:urel_error_joint} (or its simple form \eqref{ineq:urel_error}).  A quantum measurement $M$ may be called trivial, or non-informative, when it is a constant map, \textit{i.e.}, there exists a fixed probability distribution $p_{0} \in W(\Omega)$ such that $M\!\rho = p_{0}$ holds for all quantum states $\rho \in Z(\HS{H})$;  in other words, trivial measurements are the least informative among all possible measurements performed on a quantum system.  It is straightforward to confirm that the pullback and pushforward of a trivial measurement $M$ are characterised by the identity operator $\dpb{M}{\rho}f = \lev{f}{M\!\rho}$ and the constant function $\pf{M}{\rho}A = \lev{A}{\rho}$, each weighted by the expectation values of the observables concerned.  Triviality of measurement thus reduces the error \eqref{def:error_quantum-measurement} to the standard deviation $\err{A}{M}{\rho} = \lstdv{A}{\rho}$, further bringing the overall relation \eqref{ineq:urel_error_joint} towards the Schr{\"o}dinger relation \eqref{ineq:uncertainty-relation_Schroedinger}, to which the Kennard--Robertson relation is a trivial corollary.

\subsection{The Ozawa Relation}

Another notable corollary to the special case of the general relation \eqref{ineq:urel_error_joint} is the Ozawa relation \cite{Ozawa_2004_01} regarding errors of joint measurements (as well as its recent modification \cite{Ozawa_2019}).  For this, it should be noted that Ozawa's error $\varepsilon$ is defined for the special class of quantum measurements that yield real outcomes $\Omega = \mathbb{R}$, and that the error \eqref{def:error_quantum-measurement} is mathematically well-defined whenever Ozawa's error is so (in fact, the error \eqref{def:error_quantum-measurement} is never greater than Ozawa's).

Given the fact that the (global) joint measurability, upon which many formulations (including Ozawa's) are founded, is a stronger condition than that of the local joint measurability introduced in this paper, let $J$ be a (global) joint POVM measurement with outcomes in $\Omega = \mathbb{R}^{2}$.  Provided that Ozawa's errors $\epsilon(A)$ and $\epsilon(B)$ are both well-defined over the state $\rho$ for the marginal POVM measurements $M_{1}$ and $M_{2}$ of $J$, respectively, one finds the chain of inequalities $\epsilon(A)\, \epsilon(B) \geq \err{A}{M_{1}}{\rho}\, \err{B}{M_{2}}{\rho} \geq ( \mathcal{R}^2 + \mathcal{I}^2 )^{1/2} \geq \abs{ \mathcal{I} } \geq \abs{ \lev{ [A,B] }{\rho} } /2 - \epsilon(A)\,\lstdv{B}{\rho} - \lstdv{A}{\rho}\,\epsilon(B)$ with $\mathcal{R}$ and $\mathcal{I}$ being respectively the semiclassical \eqref{def:urel_joint_real} and quantum \eqref{def:urel_joint_imaginary} contributions to the lower bound of the product of the errors.  Here, note that the left- and right-most-hand sides of the above chain is equivalent to the Ozawa relation, whereas the inequality in the middle is the relation \eqref{ineq:urel_error_joint} under local joint-measurability.

\subsection{Unbiasedness of Measurement}

The notion of local representability introduced in this paper can in essence be understood to be a more general and universal concept than those of `unbiasedness' found in the operator-theoretic formulation of Arthurs--Kelly--Goodman, or those of `local unbiasedness' found in the framework of estimation theory adopted by Watanabe, Sagawa, and Ueda in formulating their relations.

\subsubsection{The Arthurs--Kelly--Goodman Relation}

The Arthurs--Kelly--Goodman relation \cite{Arthurs_1965,Arthurs_1988} is another noteworthy corollary to the special case of the framework presented in this paper.  In view of the fact that Ozawa's definition of error is an extension of AKG's, the latter of which is only relevant when the measurement fulfils the (global) `unbiasedness' condition regarding the measurement of the observables concerned, the same remarks regarding Ozawa's formulation also apply to AKG's.  Given the fact that the concept of unbiased measurement adopted by AKG is a much stronger assumption than that of local representability introduced in this paper, one confirms that both definitions \eqref{def:error_quantum-measurement} and \eqref{def:error_quantum-measurement_representability} of the error are well-defined whenever AKG's error is, and subsequently finds that the former two errors of the author is never greater than AKG's.

With the same notations and assumptions as above, it is straightforward to find the chain of inequalities $\epsilon(A)\, \epsilon(B) \geq \rerr{A}{M_{1}}{\rho}\, \rerr{B}{M_{2}}{\rho} \geq ( \tilde{\mathcal{R}}^2 + \mathcal{I}_{0}^2 )^{1/2} \geq \abs{ \lev{ [A,B] }{\rho} } /2$ with $\tilde{\mathcal{R}}$ and $\mathcal{I}_{0}$ being respectively the semiclassical \eqref{def:urel_representability_joint_real} and quantum \eqref{def:urel_imaginary_representability} contributions to the lower bound of the product of the errors.  Here, the left- and right-most-hand sides of the above chain is the AKG relation, whereas the inequality in the middle is the relation \eqref{ineq:urel_error_representability_joint} under local representability.

As for the other AKG's relation regarding the standard deviations of the measurement outcomes, one finds the chain of inequalities $\lstdv{M_{1}}{\rho}\,\lstdv{M_{2}}{\rho} \geq \lstdv{\tilde{f}_{A}}{M\!\rho}\,\lstdv{\tilde{f}_{B}}{N\!\rho} \geq ( ( \abs{\tilde{\mathcal{R}}} + \abs{\mathcal{R}_{0}} )^{2} + 4\,\mathcal{I}_{0}^{2} )^{1/2} \geq \abs{ \lev{ [A,B] }{\rho} }$ with $\tilde{f}_{A}$ and $\tilde{f}_{B}$ respectively being the optimal $\rho$-representatives of $A$ and $B$ dictated by the partial inverse of the pullbacks.  Here, $\tilde{\mathcal{R}}$, $\mathcal{R}_{0}$, and $\mathcal{I}_{0}$ are respectively given by \eqref{def:urel_representability_joint_real}, \eqref{def:quantum-covariance}, and \eqref{def:urel_imaginary_representability}.  The left- and right-most-hand sides of the above chain is the AKG relation, whereas the inequality in the middle is the relation \eqref{ineq:uncert_joint_representability_2} presented in this paper. 
 
\subsubsection{The Watanabe--Sagawa--Ueda Relation}

The Watanabe--Sagawa--Ueda relation \cite{Watanabe_2011} adopts the estimation theory \cite{Holevo_1982} as its grounding framework.  As such, they presume a differential structure on the state-spaces, which the framework of this paper does not necessitate;  in this regard, the framework of this paper is more universal than theirs  (it is also worth noting that their results are only accountable for quantum measurements on finite-dimensional Hilbert spaces, whereas the results of this paper has no such limits).  Specifically, the notion of local representability (and local representatives) introduced in this paper is found to be a more universal concept than those of local unbiasedness (and locally unbiased estimators) adopted in their framework;  the former concept may be defined without any additional structure (such as differential structure), whereas the latter cannot do without them.

It is then found that the Watanabe--Sagawa--Ueda relation for errors of quantum measurements may be understood in essence to be a corollary to the special case of the uncertainty relation \eqref{ineq:urel_error_representability_joint} under local representability.  Since a detailed discussion on this topic requires much exposure to the estimation theory (especially differential geometry), it is beyond the scope of this paper.  A comprehensive description will be given elsewhere in later publications of the author.

\section{Discussions\label{sec:discussions}}

The tools, concepts, and results introduced in this paper has been so far mostly explicated for quantum measurements, which are affine maps from quantum-state spaces to classical-state spaces (occasional comments have been also made for classical measurements as well).  In view of the fact that the methods presented here are highly universal, it is straightforward to observe that the applicability of the framework of this paper are not limited to quantum measurements, but also extends equally to any maps between state spaces (which shall be called \emph{processes} in generic terms), thereby leading to analogous concepts, definitions, and results for general processes as those presented in this paper.

Among the most interesting classes of processes within the context of quantum theory would be the following four types of maps:  those from quantum-state spaces to classical-state spaces $M : Z(\mathcal{H}) \to W(\Omega)$, those between classical-state spaces $K : W(\Omega_{1}) \to W(\Omega_{2})$, those between quantum-state spaces $\Theta : Z(\mathcal{H}) \to Z(\mathcal{K})$, and those from classical-state spaces to quantum-state spaces $\Phi : W(\Omega) \to Z(\mathcal{H})$.  In this paper, the first quantum-to-classical (Q-C) process $M$ has beed interpreted as a quantum measurement, whereas the second classical-to-classical (C-C) process $K$ has been addressed as a classical measurement (or classical process).

Application of the universal framework to general processes shall be explicated in detail in the subsequent papers of the author.  Among the most interesting are its application to quantum processes (\textit{i.e.}, quantum-to-quantum (Q-Q) processes), which notably results in the uncertainty relation for error and disturbance (see Ref.~\cite{Lee_2020_02}).  More detailed and extensive results on the relations for error and disturbance (specifically under local representability) shall be reported shortly in an upcoming work of the author.

\begin{acknowledgments}
This work was supported by JSPS Grant-in-Aid for Scientific Research (KAKENHI), Grant Numbers JP18K13468 and JP20H01906.
\end{acknowledgments}

\bibliography{urel}

\providecommand{\noopsort}[1]{}\providecommand{\singleletter}[1]{#1}%
\begin{thebibliography}{66}%
\makeatletter
\providecommand \@ifxundefined [1]{%
 \@ifx{#1\undefined}
}%
\providecommand \@ifnum [1]{%
 \ifnum #1\expandafter \@firstoftwo
 \else \expandafter \@secondoftwo
 \fi
}%
\providecommand \@ifx [1]{%
 \ifx #1\expandafter \@firstoftwo
 \else \expandafter \@secondoftwo
 \fi
}%
\providecommand \natexlab [1]{#1}%
\providecommand \enquote  [1]{``#1''}%
\providecommand \bibnamefont  [1]{#1}%
\providecommand \bibfnamefont [1]{#1}%
\providecommand \citenamefont [1]{#1}%
\providecommand \href@noop [0]{\@secondoftwo}%
\providecommand \href [0]{\begingroup \@sanitize@url \@href}%
\providecommand \@href[1]{\@@startlink{#1}\@@href}%
\providecommand \@@href[1]{\endgroup#1\@@endlink}%
\providecommand \@sanitize@url [0]{\catcode `\\12\catcode `\$12\catcode
  `\&12\catcode `\#12\catcode `\^12\catcode `\_12\catcode `\%12\relax}%
\providecommand \@@startlink[1]{}%
\providecommand \@@endlink[0]{}%
\providecommand \url  [0]{\begingroup\@sanitize@url \@url }%
\providecommand \@url [1]{\endgroup\@href {#1}{\urlprefix }}%
\providecommand \urlprefix  [0]{URL }%
\providecommand \Eprint [0]{\href }%
\providecommand \doibase [0]{https://doi.org/}%
\providecommand \selectlanguage [0]{\@gobble}%
\providecommand \bibinfo  [0]{\@secondoftwo}%
\providecommand \bibfield  [0]{\@secondoftwo}%
\providecommand \translation [1]{[#1]}%
\providecommand \BibitemOpen [0]{}%
\providecommand \bibitemStop [0]{}%
\providecommand \bibitemNoStop [0]{.\EOS\space}%
\providecommand \EOS [0]{\spacefactor3000\relax}%
\providecommand \BibitemShut  [1]{\csname bibitem#1\endcsname}%
\let\auto@bib@innerbib\@empty
\bibitem [{\citenamefont {Heisenberg}(1927)}]{Heisenberg_1927}%
  \BibitemOpen
  \bibfield  {author} {\bibinfo {author} {\bibfnamefont {W.~K.}\ \bibnamefont
  {Heisenberg}},\ }\href {https://doi.org/10.1007/BF01397280} {\bibfield
  {journal} {\bibinfo  {journal} {Z. Phys.}\ }\textbf {\bibinfo {volume}
  {43}},\ \bibinfo {pages} {172} (\bibinfo {year} {1927})}\BibitemShut
  {NoStop}%
\bibitem [{\citenamefont {Kennard}(1927)}]{Kennard_1927}%
  \BibitemOpen
  \bibfield  {author} {\bibinfo {author} {\bibfnamefont {E.~H.}\ \bibnamefont
  {Kennard}},\ }\href {https://doi.org/10.1007/BF01391200} {\bibfield
  {journal} {\bibinfo  {journal} {Z. Phys.}\ }\textbf {\bibinfo {volume}
  {44}},\ \bibinfo {pages} {326} (\bibinfo {year} {1927})}\BibitemShut
  {NoStop}%
\bibitem [{\citenamefont {Robertson}(1929)}]{Robertson_1929}%
  \BibitemOpen
  \bibfield  {author} {\bibinfo {author} {\bibfnamefont {H.~P.}\ \bibnamefont
  {Robertson}},\ }\href {https://doi.org/10.1103/PhysRev.34.163} {\bibfield
  {journal} {\bibinfo  {journal} {Phys. Rev.}\ }\textbf {\bibinfo {volume}
  {34}},\ \bibinfo {pages} {163} (\bibinfo {year} {1929})}\BibitemShut
  {NoStop}%
\bibitem [{\citenamefont {Heisenberg}(1930)}]{Heisenberg_1930}%
  \BibitemOpen
  \bibfield  {author} {\bibinfo {author} {\bibfnamefont {W.~K.}\ \bibnamefont
  {Heisenberg}},\ }\href@noop {} {\emph {\bibinfo {title} {{D}ie physikalischen
  {P}rinzipien der {Q}uantentheorie}}}\ (\bibinfo  {publisher} {S. Hirzel
  Verlag},\ \bibinfo {address} {Leipzig},\ \bibinfo {year} {1930})\BibitemShut
  {NoStop}%
\bibitem [{\citenamefont {Arthurs}\ and\ \citenamefont
  {Kelly~Jr.}(1965)}]{Arthurs_1965}%
  \BibitemOpen
  \bibfield  {author} {\bibinfo {author} {\bibfnamefont {E.}~\bibnamefont
  {Arthurs}}\ and\ \bibinfo {author} {\bibfnamefont {J.~L.}\ \bibnamefont
  {Kelly~Jr.}},\ }\href {https://doi.org/10.1002/j.1538-7305.1965.tb01684.x}
  {\bibfield  {journal} {\bibinfo  {journal} {Bell Sys. Tech. J.}\ }\textbf
  {\bibinfo {volume} {44}},\ \bibinfo {pages} {725} (\bibinfo {year}
  {1965})}\BibitemShut {NoStop}%
\bibitem [{\citenamefont {Arthurs}\ and\ \citenamefont
  {Goodman}(1988)}]{Arthurs_1988}%
  \BibitemOpen
  \bibfield  {author} {\bibinfo {author} {\bibfnamefont {E.}~\bibnamefont
  {Arthurs}}\ and\ \bibinfo {author} {\bibfnamefont {M.~S.}\ \bibnamefont
  {Goodman}},\ }\href {https://doi.org/10.1103/PhysRevLett.60.2447} {\bibfield
  {journal} {\bibinfo  {journal} {Phys. Rev. Lett.}\ }\textbf {\bibinfo
  {volume} {60}},\ \bibinfo {pages} {2447} (\bibinfo {year}
  {1988})}\BibitemShut {NoStop}%
\bibitem [{\citenamefont {Ozawa}(2003)}]{Ozawa_2003}%
  \BibitemOpen
  \bibfield  {author} {\bibinfo {author} {\bibfnamefont {M.}~\bibnamefont
  {Ozawa}},\ }\href {https://doi.org/10.1103/PhysRevA.67.042105} {\bibfield
  {journal} {\bibinfo  {journal} {Phys. Rev. A}\ }\textbf {\bibinfo {volume}
  {67}},\ \bibinfo {pages} {042105} (\bibinfo {year} {2003})}\BibitemShut
  {NoStop}%
\bibitem [{\citenamefont {Ozawa}(2004)}]{Ozawa_2004_01}%
  \BibitemOpen
  \bibfield  {author} {\bibinfo {author} {\bibfnamefont {M.}~\bibnamefont
  {Ozawa}},\ }\href {https://doi.org/10.1016/j.physleta.2003.12.001} {\bibfield
   {journal} {\bibinfo  {journal} {Phys. Lett. A}\ }\textbf {\bibinfo {volume}
  {320}},\ \bibinfo {pages} {367} (\bibinfo {year} {2004})}\BibitemShut
  {NoStop}%
\bibitem [{\citenamefont {Branciard}(2013)}]{Branciard_2013}%
  \BibitemOpen
  \bibfield  {author} {\bibinfo {author} {\bibfnamefont {C.}~\bibnamefont
  {Branciard}},\ }\href {https://doi.org/10.1073/pnas.1219331110} {\bibfield
  {journal} {\bibinfo  {journal} {PNAS}\ }\textbf {\bibinfo {volume} {110}},\
  \bibinfo {pages} {6742} (\bibinfo {year} {2013})}\BibitemShut {NoStop}%
\bibitem [{\citenamefont {Ozawa}(2019)}]{Ozawa_2019}%
  \BibitemOpen
  \bibfield  {author} {\bibinfo {author} {\bibfnamefont {M.}~\bibnamefont
  {Ozawa}},\ }\href {https://doi.org/10.1038/s41534-018-0113-z} {\bibfield
  {journal} {\bibinfo  {journal} {npj Quant. Inf.}\ }\textbf {\bibinfo {volume}
  {5}},\ \bibinfo {pages} {1} (\bibinfo {year} {2019})}\BibitemShut {NoStop}%
\bibitem [{\citenamefont {Werner}(2004)}]{Werner_2004}%
  \BibitemOpen
  \bibfield  {author} {\bibinfo {author} {\bibfnamefont {R.~F.}\ \bibnamefont
  {Werner}},\ }\href@noop {} {\bibfield  {journal} {\bibinfo  {journal} {Quant.
  Inf. Comput.}\ }\textbf {\bibinfo {volume} {4}},\ \bibinfo {pages} {546}
  (\bibinfo {year} {2004})}\BibitemShut {NoStop}%
\bibitem [{\citenamefont {Miyadera}\ and\ \citenamefont
  {Imai}(2008)}]{Miyadera_2008}%
  \BibitemOpen
  \bibfield  {author} {\bibinfo {author} {\bibfnamefont {T.}~\bibnamefont
  {Miyadera}}\ and\ \bibinfo {author} {\bibfnamefont {H.}~\bibnamefont
  {Imai}},\ }\href {https://doi.org/10.1103/PhysRevA.78.052119} {\bibfield
  {journal} {\bibinfo  {journal} {Phys. Rev. A}\ }\textbf {\bibinfo {volume}
  {78}},\ \bibinfo {pages} {052119} (\bibinfo {year} {2008})}\BibitemShut
  {NoStop}%
\bibitem [{\citenamefont {Busch}\ \emph {et~al.}(2013)\citenamefont {Busch},
  \citenamefont {Lahti},\ and\ \citenamefont {Werner}}]{Busch_2013}%
  \BibitemOpen
  \bibfield  {author} {\bibinfo {author} {\bibfnamefont {P.}~\bibnamefont
  {Busch}}, \bibinfo {author} {\bibfnamefont {P.}~\bibnamefont {Lahti}},\ and\
  \bibinfo {author} {\bibfnamefont {R.~F.}\ \bibnamefont {Werner}},\ }\href
  {https://doi.org/10.1103/PhysRevLett.111.160405} {\bibfield  {journal}
  {\bibinfo  {journal} {Phys. Rev. Lett.}\ }\textbf {\bibinfo {volume} {111}},\
  \bibinfo {pages} {160405} (\bibinfo {year} {2013})}\BibitemShut {NoStop}%
\bibitem [{\citenamefont {Yuen}\ and\ \citenamefont {Lax}(1973)}]{Yuen_1973}%
  \BibitemOpen
  \bibfield  {author} {\bibinfo {author} {\bibfnamefont {H.}~\bibnamefont
  {Yuen}}\ and\ \bibinfo {author} {\bibfnamefont {M.}~\bibnamefont {Lax}},\
  }\href {https://doi.org/10.1109/TIT.1973.1055103} {\bibfield  {journal}
  {\bibinfo  {journal} {IEEE Transactions on Information Theory}\ }\textbf
  {\bibinfo {volume} {19}},\ \bibinfo {pages} {740} (\bibinfo {year}
  {1973})}\BibitemShut {NoStop}%
\bibitem [{\citenamefont {Watanabe}\ \emph {et~al.}(2011)\citenamefont
  {Watanabe}, \citenamefont {Sagawa},\ and\ \citenamefont
  {Ueda}}]{Watanabe_2011}%
  \BibitemOpen
  \bibfield  {author} {\bibinfo {author} {\bibfnamefont {Y.}~\bibnamefont
  {Watanabe}}, \bibinfo {author} {\bibfnamefont {T.}~\bibnamefont {Sagawa}},\
  and\ \bibinfo {author} {\bibfnamefont {M.}~\bibnamefont {Ueda}},\ }\href
  {https://doi.org/10.1103/PhysRevA.84.042121} {\bibfield  {journal} {\bibinfo
  {journal} {Phys. Rev. A}\ }\textbf {\bibinfo {volume} {84}},\ \bibinfo
  {pages} {042121} (\bibinfo {year} {2011})}\BibitemShut {NoStop}%
\bibitem [{\citenamefont {Mandelshtam}\ and\ \citenamefont
  {Tamm}(1945)}]{Mandelshtam_1945}%
  \BibitemOpen
  \bibfield  {author} {\bibinfo {author} {\bibfnamefont {L.~I.}\ \bibnamefont
  {Mandelshtam}}\ and\ \bibinfo {author} {\bibfnamefont {I.~E.}\ \bibnamefont
  {Tamm}},\ }\href@noop {} {\bibfield  {journal} {\bibinfo  {journal} {Izv.
  Akad. Nauk SSSR, Ser. Fiz.}\ }\textbf {\bibinfo {volume} {9}},\ \bibinfo
  {pages} {122} (\bibinfo {year} {1945})}\BibitemShut {NoStop}%
\bibitem [{\citenamefont {Allcock}(1969{\natexlab{a}})}]{Allcock_1969_01}%
  \BibitemOpen
  \bibfield  {author} {\bibinfo {author} {\bibfnamefont {G.~R.}\ \bibnamefont
  {Allcock}},\ }\href {https://doi.org/10.1016/0003-4916(69)90251-6} {\bibfield
   {journal} {\bibinfo  {journal} {Ann. Phys.}\ }\textbf {\bibinfo {volume}
  {53}},\ \bibinfo {pages} {253} (\bibinfo {year}
  {1969}{\natexlab{a}})}\BibitemShut {NoStop}%
\bibitem [{\citenamefont {Allcock}(1969{\natexlab{b}})}]{Allcock_1969_02}%
  \BibitemOpen
  \bibfield  {author} {\bibinfo {author} {\bibfnamefont {G.~R.}\ \bibnamefont
  {Allcock}},\ }\href {https://doi.org/10.1016/0003-4916(69)90252-8} {\bibfield
   {journal} {\bibinfo  {journal} {Ann. Phys.}\ }\textbf {\bibinfo {volume}
  {53}},\ \bibinfo {pages} {286} (\bibinfo {year}
  {1969}{\natexlab{b}})}\BibitemShut {NoStop}%
\bibitem [{\citenamefont {Allcock}(1969{\natexlab{c}})}]{Allcock_1969_03}%
  \BibitemOpen
  \bibfield  {author} {\bibinfo {author} {\bibfnamefont {G.~R.}\ \bibnamefont
  {Allcock}},\ }\href {https://doi.org/10.1016/0003-4916(69)90253-X} {\bibfield
   {journal} {\bibinfo  {journal} {Ann. Phys.}\ }\textbf {\bibinfo {volume}
  {53}},\ \bibinfo {pages} {311} (\bibinfo {year}
  {1969}{\natexlab{c}})}\BibitemShut {NoStop}%
\bibitem [{\citenamefont {Helstrom}(1976)}]{Helstrom_1976}%
  \BibitemOpen
  \bibfield  {author} {\bibinfo {author} {\bibfnamefont {C.~W.}\ \bibnamefont
  {Helstrom}},\ }\href@noop {} {\emph {\bibinfo {title} {Quantum Detection and
  Estimation Theory}}}\ (\bibinfo  {publisher} {Academic Press},\ \bibinfo
  {address} {Cambridge, MA, USA},\ \bibinfo {year} {1976})\BibitemShut
  {NoStop}%
\bibitem [{\citenamefont {Hirschman~Jr.}(1957)}]{Hirschman_1957}%
  \BibitemOpen
  \bibfield  {author} {\bibinfo {author} {\bibfnamefont {I.~I.}\ \bibnamefont
  {Hirschman~Jr.}},\ }\href {https://doi.org/10.2307/2372390} {\bibfield
  {journal} {\bibinfo  {journal} {Am. J. Math.}\ }\textbf {\bibinfo {volume}
  {79}},\ \bibinfo {pages} {152} (\bibinfo {year} {1957})}\BibitemShut
  {NoStop}%
\bibitem [{\citenamefont {Beckner}(1975)}]{Beckner_1975}%
  \BibitemOpen
  \bibfield  {author} {\bibinfo {author} {\bibfnamefont {W.}~\bibnamefont
  {Beckner}},\ }\href {https://doi.org/10.2307/1970980} {\bibfield  {journal}
  {\bibinfo  {journal} {Ann. Math.}\ }\textbf {\bibinfo {volume} {102}},\
  \bibinfo {pages} {159} (\bibinfo {year} {1975})}\BibitemShut {NoStop}%
\bibitem [{\citenamefont {Bia{\l}ynicki-Birula}\ and\ \citenamefont
  {Mycielski}(1975)}]{Birula_1975}%
  \BibitemOpen
  \bibfield  {author} {\bibinfo {author} {\bibfnamefont {I.}~\bibnamefont
  {Bia{\l}ynicki-Birula}}\ and\ \bibinfo {author} {\bibfnamefont
  {J.}~\bibnamefont {Mycielski}},\ }\href {https://doi.org/10.1007/BF01608825}
  {\bibfield  {journal} {\bibinfo  {journal} {Commun. Math. Phys.}\ }\textbf
  {\bibinfo {volume} {44}},\ \bibinfo {pages} {129} (\bibinfo {year}
  {1975})}\BibitemShut {NoStop}%
\bibitem [{\citenamefont {Deutsch}(1983)}]{Deutsch_1983}%
  \BibitemOpen
  \bibfield  {author} {\bibinfo {author} {\bibfnamefont {D.}~\bibnamefont
  {Deutsch}},\ }\href {https://doi.org/10.1103/PhysRevLett.50.631} {\bibfield
  {journal} {\bibinfo  {journal} {Phys. Rev. Lett.}\ }\textbf {\bibinfo
  {volume} {50}},\ \bibinfo {pages} {631} (\bibinfo {year} {1983})}\BibitemShut
  {NoStop}%
\bibitem [{\citenamefont {Wigner}(1952)}]{Wigner_1952}%
  \BibitemOpen
  \bibfield  {author} {\bibinfo {author} {\bibfnamefont {E.~P.}\ \bibnamefont
  {Wigner}},\ }\href {https://doi.org/10.1007/BF01948686} {\bibfield  {journal}
  {\bibinfo  {journal} {Z. Phys.}\ }\textbf {\bibinfo {volume} {133}},\
  \bibinfo {pages} {101} (\bibinfo {year} {1952})}\BibitemShut {NoStop}%
\bibitem [{\citenamefont {Araki}\ and\ \citenamefont
  {Yanase}(1960)}]{Araki_1960}%
  \BibitemOpen
  \bibfield  {author} {\bibinfo {author} {\bibfnamefont {H.}~\bibnamefont
  {Araki}}\ and\ \bibinfo {author} {\bibfnamefont {M.~M.}\ \bibnamefont
  {Yanase}},\ }\href {https://doi.org/10.1103/PhysRev.120.622} {\bibfield
  {journal} {\bibinfo  {journal} {Phys. Rev.}\ }\textbf {\bibinfo {volume}
  {120}},\ \bibinfo {pages} {622} (\bibinfo {year} {1960})}\BibitemShut
  {NoStop}%
\bibitem [{\citenamefont {Yanase}(1961)}]{Yanase_1961}%
  \BibitemOpen
  \bibfield  {author} {\bibinfo {author} {\bibfnamefont {M.~M.}\ \bibnamefont
  {Yanase}},\ }\href {https://doi.org/10.1103/PhysRev.123.666} {\bibfield
  {journal} {\bibinfo  {journal} {Phys. Rev.}\ }\textbf {\bibinfo {volume}
  {123}},\ \bibinfo {pages} {666} (\bibinfo {year} {1961})}\BibitemShut
  {NoStop}%
\bibitem [{\citenamefont {Ozawa}(2002{\natexlab{a}})}]{Ozawa_2002_01}%
  \BibitemOpen
  \bibfield  {author} {\bibinfo {author} {\bibfnamefont {M.}~\bibnamefont
  {Ozawa}},\ }\href {https://doi.org/10.1103/PhysRevLett.88.050402} {\bibfield
  {journal} {\bibinfo  {journal} {Phys. Rev. Lett.}\ }\textbf {\bibinfo
  {volume} {88}},\ \bibinfo {pages} {050402} (\bibinfo {year}
  {2002}{\natexlab{a}})}\BibitemShut {NoStop}%
\bibitem [{\citenamefont {Fleming}(1973)}]{Fleming_1973}%
  \BibitemOpen
  \bibfield  {author} {\bibinfo {author} {\bibfnamefont {G.~N.}\ \bibnamefont
  {Fleming}},\ }\href {https://doi.org/10.1007/BF02819419} {\bibfield
  {journal} {\bibinfo  {journal} {Nuov. Cim. A}\ }\textbf {\bibinfo {volume}
  {16}},\ \bibinfo {pages} {232} (\bibinfo {year} {1973})}\BibitemShut
  {NoStop}%
\bibitem [{\citenamefont {Anandan}\ and\ \citenamefont
  {Aharonov}(1990)}]{Aharonov_1990}%
  \BibitemOpen
  \bibfield  {author} {\bibinfo {author} {\bibfnamefont {J.}~\bibnamefont
  {Anandan}}\ and\ \bibinfo {author} {\bibfnamefont {Y.}~\bibnamefont
  {Aharonov}},\ }\href {https://doi.org/10.1103/PhysRevLett.65.1697} {\bibfield
   {journal} {\bibinfo  {journal} {Phys. Rev. Lett.}\ }\textbf {\bibinfo
  {volume} {65}},\ \bibinfo {pages} {1697} (\bibinfo {year}
  {1990})}\BibitemShut {NoStop}%
\bibitem [{\citenamefont {Pfeifer}(1993)}]{Pfeifer_1993}%
  \BibitemOpen
  \bibfield  {author} {\bibinfo {author} {\bibfnamefont {P.}~\bibnamefont
  {Pfeifer}},\ }\href {https://doi.org/10.1103/PhysRevLett.70.3365} {\bibfield
  {journal} {\bibinfo  {journal} {Phys. Rev. Lett.}\ }\textbf {\bibinfo
  {volume} {70}},\ \bibinfo {pages} {3365} (\bibinfo {year}
  {1993})}\BibitemShut {NoStop}%
\bibitem [{\citenamefont {Margolus}\ and\ \citenamefont
  {Levitin}(1998)}]{Margolus_1998}%
  \BibitemOpen
  \bibfield  {author} {\bibinfo {author} {\bibfnamefont {N.}~\bibnamefont
  {Margolus}}\ and\ \bibinfo {author} {\bibfnamefont {L.~B.}\ \bibnamefont
  {Levitin}},\ }\href {https://doi.org/10.1016/S0167-2789(98)00054-2}
  {\bibfield  {journal} {\bibinfo  {journal} {Physica D: Nonlinear Phenomena}\
  }\textbf {\bibinfo {volume} {120}},\ \bibinfo {pages} {188} (\bibinfo {year}
  {1998})}\BibitemShut {NoStop}%
\bibitem [{\citenamefont {Giovannetti}\ \emph {et~al.}(2003)\citenamefont
  {Giovannetti}, \citenamefont {Lloyd},\ and\ \citenamefont
  {Maccone}}]{Giovannetti_2003}%
  \BibitemOpen
  \bibfield  {author} {\bibinfo {author} {\bibfnamefont {V.}~\bibnamefont
  {Giovannetti}}, \bibinfo {author} {\bibfnamefont {S.}~\bibnamefont {Lloyd}},\
  and\ \bibinfo {author} {\bibfnamefont {L.}~\bibnamefont {Maccone}},\ }\href
  {https://doi.org/10.1103/PhysRevA.67.052109} {\bibfield  {journal} {\bibinfo
  {journal} {Phys. Rev. A}\ }\textbf {\bibinfo {volume} {67}},\ \bibinfo
  {pages} {052109} (\bibinfo {year} {2003})}\BibitemShut {NoStop}%
\bibitem [{\citenamefont {Jones}\ and\ \citenamefont {Kok}(2010)}]{Jones_2010}%
  \BibitemOpen
  \bibfield  {author} {\bibinfo {author} {\bibfnamefont {P.~J.}\ \bibnamefont
  {Jones}}\ and\ \bibinfo {author} {\bibfnamefont {P.}~\bibnamefont {Kok}},\
  }\href {https://doi.org/10.1103/PhysRevA.82.022107} {\bibfield  {journal}
  {\bibinfo  {journal} {Phys. Rev. A}\ }\textbf {\bibinfo {volume} {82}},\
  \bibinfo {pages} {022107} (\bibinfo {year} {2010})}\BibitemShut {NoStop}%
\bibitem [{\citenamefont {Pires}\ \emph {et~al.}(2016)\citenamefont {Pires},
  \citenamefont {Cianciaruso}, \citenamefont {C{\'e}leri}, \citenamefont
  {Adesso},\ and\ \citenamefont {Soares-Pinto}}]{Pires_2016}%
  \BibitemOpen
  \bibfield  {author} {\bibinfo {author} {\bibfnamefont {D.~P.}\ \bibnamefont
  {Pires}}, \bibinfo {author} {\bibfnamefont {M.}~\bibnamefont {Cianciaruso}},
  \bibinfo {author} {\bibfnamefont {L.~C.}\ \bibnamefont {C{\'e}leri}},
  \bibinfo {author} {\bibfnamefont {G.}~\bibnamefont {Adesso}},\ and\ \bibinfo
  {author} {\bibfnamefont {D.~O.}\ \bibnamefont {Soares-Pinto}},\ }\href
  {https://doi.org/10.1103/PhysRevX.6.021031} {\bibfield  {journal} {\bibinfo
  {journal} {Phys. Rev. X}\ }\textbf {\bibinfo {volume} {6}},\ \bibinfo {pages}
  {021031} (\bibinfo {year} {2016})}\BibitemShut {NoStop}%
\bibitem [{\citenamefont {Shiraishi}\ \emph {et~al.}(2018)\citenamefont
  {Shiraishi}, \citenamefont {Funo},\ and\ \citenamefont
  {Saito}}]{Shiraishi_2018}%
  \BibitemOpen
  \bibfield  {author} {\bibinfo {author} {\bibfnamefont {N.}~\bibnamefont
  {Shiraishi}}, \bibinfo {author} {\bibfnamefont {K.}~\bibnamefont {Funo}},\
  and\ \bibinfo {author} {\bibfnamefont {K.}~\bibnamefont {Saito}},\ }\href
  {https://doi.org/10.1103/PhysRevLett.121.070601} {\bibfield  {journal}
  {\bibinfo  {journal} {Phys. Rev. Lett.}\ }\textbf {\bibinfo {volume} {121}},\
  \bibinfo {pages} {070601} (\bibinfo {year} {2018})}\BibitemShut {NoStop}%
\bibitem [{\citenamefont {Ozawa}(2002{\natexlab{b}})}]{Ozawa_2002_02}%
  \BibitemOpen
  \bibfield  {author} {\bibinfo {author} {\bibfnamefont {M.}~\bibnamefont
  {Ozawa}},\ }\href {https://doi.org/10.1103/PhysRevLett.89.057902} {\bibfield
  {journal} {\bibinfo  {journal} {Phys. Rev. Lett.}\ }\textbf {\bibinfo
  {volume} {89}},\ \bibinfo {pages} {057902} (\bibinfo {year}
  {2002}{\natexlab{b}})}\BibitemShut {NoStop}%
\bibitem [{\citenamefont {Tajima}\ \emph {et~al.}(2018)\citenamefont {Tajima},
  \citenamefont {Shiraishi},\ and\ \citenamefont {Saito}}]{Tajima_2018}%
  \BibitemOpen
  \bibfield  {author} {\bibinfo {author} {\bibfnamefont {H.}~\bibnamefont
  {Tajima}}, \bibinfo {author} {\bibfnamefont {N.}~\bibnamefont {Shiraishi}},\
  and\ \bibinfo {author} {\bibfnamefont {K.}~\bibnamefont {Saito}},\ }\href
  {https://doi.org/10.1103/PhysRevLett.121.110403} {\bibfield  {journal}
  {\bibinfo  {journal} {Phys. Rev. Lett.}\ }\textbf {\bibinfo {volume} {121}},\
  \bibinfo {pages} {110403} (\bibinfo {year} {2018})}\BibitemShut {NoStop}%
\bibitem [{\citenamefont {Hall}(2001)}]{Hall_2001}%
  \BibitemOpen
  \bibfield  {author} {\bibinfo {author} {\bibfnamefont {M.~J.~W.}\
  \bibnamefont {Hall}},\ }\href {https://doi.org/10.1103/PhysRevA.64.052103}
  {\bibfield  {journal} {\bibinfo  {journal} {Phys. Rev. A}\ }\textbf {\bibinfo
  {volume} {64}},\ \bibinfo {pages} {052103} (\bibinfo {year}
  {2001})}\BibitemShut {NoStop}%
\bibitem [{\citenamefont {Dressel}\ and\ \citenamefont
  {Nori}(2014)}]{Dressel_2014}%
  \BibitemOpen
  \bibfield  {author} {\bibinfo {author} {\bibfnamefont {J.}~\bibnamefont
  {Dressel}}\ and\ \bibinfo {author} {\bibfnamefont {F.}~\bibnamefont {Nori}},\
  }\href {https://doi.org/10.1103/PhysRevA.89.022106} {\bibfield  {journal}
  {\bibinfo  {journal} {Phys. Rev. A}\ }\textbf {\bibinfo {volume} {89}},\
  \bibinfo {pages} {022106} (\bibinfo {year} {2014})}\BibitemShut {NoStop}%
\bibitem [{\citenamefont {Lee}\ and\ \citenamefont {Tsutsui}(2016)}]{Lee_2016}%
  \BibitemOpen
  \bibfield  {author} {\bibinfo {author} {\bibfnamefont {J.}~\bibnamefont
  {Lee}}\ and\ \bibinfo {author} {\bibfnamefont {I.}~\bibnamefont {Tsutsui}},\
  }\href {https://doi.org/10.1016/j.physleta.2016.04.009} {\bibfield  {journal}
  {\bibinfo  {journal} {Phys. Lett. A}\ }\textbf {\bibinfo {volume} {380}},\
  \bibinfo {pages} {2045} (\bibinfo {year} {2016})}\BibitemShut {NoStop}%
\bibitem [{\citenamefont {Pollak}\ and\ \citenamefont
  {Miret-Art{\'e}s}(2019)}]{Pollak_2019}%
  \BibitemOpen
  \bibfield  {author} {\bibinfo {author} {\bibfnamefont {E.}~\bibnamefont
  {Pollak}}\ and\ \bibinfo {author} {\bibfnamefont {S.}~\bibnamefont
  {Miret-Art{\'e}s}},\ }\href {https://doi.org/10.1103/PhysRevA.99.012108}
  {\bibfield  {journal} {\bibinfo  {journal} {Phys. Rev. A}\ }\textbf {\bibinfo
  {volume} {99}},\ \bibinfo {pages} {012108} (\bibinfo {year}
  {2019})}\BibitemShut {NoStop}%
\bibitem [{\citenamefont {Koshino}\ and\ \citenamefont
  {Shimizu}(2005)}]{Koshino_2005}%
  \BibitemOpen
  \bibfield  {author} {\bibinfo {author} {\bibfnamefont {K.}~\bibnamefont
  {Koshino}}\ and\ \bibinfo {author} {\bibfnamefont {A.}~\bibnamefont
  {Shimizu}},\ }\href {https://doi.org/10.1016/j.physrep.2005.03.001}
  {\bibfield  {journal} {\bibinfo  {journal} {Phys. Rep.}\ }\textbf {\bibinfo
  {volume} {412}},\ \bibinfo {pages} {191} (\bibinfo {year}
  {2005})}\BibitemShut {NoStop}%
\bibitem [{\citenamefont {Lee}\ and\ \citenamefont
  {Tsutsui}(2020{\natexlab{a}})}]{Lee_2020_01_preprint}%
  \BibitemOpen
  \bibfield  {author} {\bibinfo {author} {\bibfnamefont {J.}~\bibnamefont
  {Lee}}\ and\ \bibinfo {author} {\bibfnamefont {I.}~\bibnamefont {Tsutsui}},\
  }\href@noop {} {\bibfield  {journal} {\bibinfo  {journal} {arXiv:2002.04008}\
  } (\bibinfo {year} {2020}{\natexlab{a}})}\BibitemShut {NoStop}%
\bibitem [{\citenamefont {Lee}\ and\ \citenamefont
  {Tsutsui}(2020{\natexlab{b}})}]{Lee_2020_01_Entropy}%
  \BibitemOpen
  \bibfield  {author} {\bibinfo {author} {\bibfnamefont {J.}~\bibnamefont
  {Lee}}\ and\ \bibinfo {author} {\bibfnamefont {I.}~\bibnamefont {Tsutsui}},\
  }\href {https://doi.org/10.3390/e22111222} {\bibfield  {journal} {\bibinfo
  {journal} {Entropy}\ }\textbf {\bibinfo {volume} {22}},\ \bibinfo {pages}
  {1222} (\bibinfo {year} {2020}{\natexlab{b}})}\BibitemShut {NoStop}%
\bibitem [{\citenamefont {Born}(1926)}]{Born_1926}%
  \BibitemOpen
  \bibfield  {author} {\bibinfo {author} {\bibfnamefont {M.}~\bibnamefont
  {Born}},\ }\href {https://doi.org/10.1007/BF01397477} {\bibfield  {journal}
  {\bibinfo  {journal} {Z. Phys.}\ }\textbf {\bibinfo {volume} {37}},\ \bibinfo
  {pages} {863} (\bibinfo {year} {1926})}\BibitemShut {NoStop}%
\bibitem [{\citenamefont {Kolmogorov}(1933)}]{Kolmogorov_1933}%
  \BibitemOpen
  \bibfield  {author} {\bibinfo {author} {\bibfnamefont {A.~N.}\ \bibnamefont
  {Kolmogorov}},\ }\href {https://doi.org/10.1007/978-3-642-49888-6} {\emph
  {\bibinfo {title} {Grundbegriffe der Wahrscheinlichkeitsrechnung}}}\
  (\bibinfo  {publisher} {Springer Verlag},\ \bibinfo {address}
  {Berlin/Heidelberg, Germany},\ \bibinfo {year} {1933})\BibitemShut {NoStop}%
\bibitem [{\citenamefont {Holevo}(2001)}]{Holevo_2001}%
  \BibitemOpen
  \bibfield  {author} {\bibinfo {author} {\bibfnamefont {A.~S.}\ \bibnamefont
  {Holevo}},\ }\href {https://doi.org/10.1007/3-540-44998-1} {\emph {\bibinfo
  {title} {Statistical Structure of Quantum Theory}}},\ \bibinfo {series}
  {Lecture Notes in Physics Monographs}, Vol.~\bibinfo {volume} {67}\ (\bibinfo
   {publisher} {Springer Verlag},\ \bibinfo {address} {Berlin/Heidelberg,
  Germany},\ \bibinfo {year} {2001})\BibitemShut {NoStop}%
\bibitem [{\citenamefont {Kadison}(1952)}]{Kadison_1952}%
  \BibitemOpen
  \bibfield  {author} {\bibinfo {author} {\bibfnamefont {R.~V.}\ \bibnamefont
  {Kadison}},\ }\href {https://doi.org/10.2307/1969657} {\bibfield  {journal}
  {\bibinfo  {journal} {Ann. Math.}\ }\textbf {\bibinfo {volume} {56}},\
  \bibinfo {pages} {494} (\bibinfo {year} {1952})}\BibitemShut {NoStop}%
\bibitem [{\citenamefont {Naimark}(1940)}]{Naimark_1940}%
  \BibitemOpen
  \bibfield  {author} {\bibinfo {author} {\bibfnamefont {M.~A.}\ \bibnamefont
  {Naimark}},\ }\href@noop {} {\bibfield  {journal} {\bibinfo  {journal} {Izv.
  Acad. Nauk SSSR Ser. Mat.}\ }\textbf {\bibinfo {volume} {4}},\ \bibinfo
  {pages} {277} (\bibinfo {year} {1940})}\BibitemShut {NoStop}%
\bibitem [{\citenamefont {Naimark}(1943)}]{Naimark_1943}%
  \BibitemOpen
  \bibfield  {author} {\bibinfo {author} {\bibfnamefont {M.~A.}\ \bibnamefont
  {Naimark}},\ }\href@noop {} {\bibfield  {journal} {\bibinfo  {journal} {C. R.
  Acad. Sci. URSS}\ }\textbf {\bibinfo {volume} {41}},\ \bibinfo {pages} {359}
  (\bibinfo {year} {1943})}\BibitemShut {NoStop}%
\bibitem [{\citenamefont {Radon}(1913)}]{Radon_1913}%
  \BibitemOpen
  \bibfield  {author} {\bibinfo {author} {\bibfnamefont {J.~K.~A.}\
  \bibnamefont {Radon}},\ }\href@noop {} {\bibfield  {journal} {\bibinfo
  {journal} {Sitzungsber. der Kaiserl. Akad. der Wiss. in Wien,
  Math.-Naturewiss. Kl.}\ }\textbf {\bibinfo {volume} {122}},\ \bibinfo {pages}
  {1295} (\bibinfo {year} {1913})}\BibitemShut {NoStop}%
\bibitem [{\citenamefont {Nikod{\'y}m}(1930)}]{Nikodym_1930}%
  \BibitemOpen
  \bibfield  {author} {\bibinfo {author} {\bibfnamefont {O.~M.}\ \bibnamefont
  {Nikod{\'y}m}},\ }\href {https://doi.org/10.4064/fm-15-1-131-179} {\bibfield
  {journal} {\bibinfo  {journal} {Fundam. Math.}\ }\textbf {\bibinfo {volume}
  {15}},\ \bibinfo {pages} {131} (\bibinfo {year} {1930})}\BibitemShut
  {NoStop}%
\bibitem [{\citenamefont {Lee}\ and\ \citenamefont {Tsutsui}(2017)}]{Lee_2017}%
  \BibitemOpen
  \bibfield  {author} {\bibinfo {author} {\bibfnamefont {J.}~\bibnamefont
  {Lee}}\ and\ \bibinfo {author} {\bibfnamefont {I.}~\bibnamefont {Tsutsui}},\
  }\href {https://doi.org/10.1093/ptep/ptx024} {\bibfield  {journal} {\bibinfo
  {journal} {PTEP}\ }\textbf {\bibinfo {volume} {2017}},\ \bibinfo {pages}
  {052A01} (\bibinfo {year} {2017})}\BibitemShut {NoStop}%
\bibitem [{\citenamefont {Lee}\ and\ \citenamefont {Tsutsui}(2018)}]{Lee_2018}%
  \BibitemOpen
  \bibfield  {author} {\bibinfo {author} {\bibfnamefont {J.}~\bibnamefont
  {Lee}}\ and\ \bibinfo {author} {\bibfnamefont {I.}~\bibnamefont {Tsutsui}},\
  }in\ \href {https://doi.org/10.1007/978-981-13-2487-1\_9} {\emph {\bibinfo
  {booktitle} {{R}eality and {M}easurement in {A}lgebraic {Q}uantum
  {T}heory}}},\ \bibinfo {series} {Springer Proceedings in Mathematics and
  Statistics}, Vol.\ \bibinfo {volume} {261},\ \bibinfo {editor} {edited by\
  \bibinfo {editor} {\bibfnamefont {M.}~\bibnamefont {Ozawa}}, \bibinfo
  {editor} {\bibfnamefont {J.}~\bibnamefont {Butterfield}}, \bibinfo {editor}
  {\bibfnamefont {H.}~\bibnamefont {Halvorson}}, \bibinfo {editor}
  {\bibfnamefont {M.}~\bibnamefont {R{\'e}dei}}, \bibinfo {editor}
  {\bibfnamefont {Y.}~\bibnamefont {Kitajima}},\ and\ \bibinfo {editor}
  {\bibfnamefont {F.}~\bibnamefont {Buscemi}}}\ (\bibinfo  {publisher}
  {Springer Singapore},\ \bibinfo {year} {2018})\ pp.\ \bibinfo {pages}
  {195--228}\BibitemShut {NoStop}%
\bibitem [{\citenamefont {Fredholm}(1903)}]{Fredholm_1903}%
  \BibitemOpen
  \bibfield  {author} {\bibinfo {author} {\bibfnamefont {I.}~\bibnamefont
  {Fredholm}},\ }\href {https://doi.org/10.1007/BF02421317} {\bibfield
  {journal} {\bibinfo  {journal} {Acta Mathematica}\ }\textbf {\bibinfo
  {volume} {27}},\ \bibinfo {pages} {365} (\bibinfo {year} {1903})}\BibitemShut
  {NoStop}%
\bibitem [{\citenamefont {Hilbert}(1904{\natexlab{a}})}]{Hilbert_1904_01}%
  \BibitemOpen
  \bibfield  {author} {\bibinfo {author} {\bibfnamefont {D.}~\bibnamefont
  {Hilbert}},\ }\href@noop {} {\bibfield  {journal} {\bibinfo  {journal}
  {Nachr. Ges. Wiss. G{\"o}ttingen, Math.-Phys. Kl.}\ }\textbf {\bibinfo
  {volume} {1}},\ \bibinfo {pages} {49} (\bibinfo {year}
  {1904}{\natexlab{a}})}\BibitemShut {NoStop}%
\bibitem [{\citenamefont {Hilbert}(1904{\natexlab{b}})}]{Hilbert_1904_02}%
  \BibitemOpen
  \bibfield  {author} {\bibinfo {author} {\bibfnamefont {D.}~\bibnamefont
  {Hilbert}},\ }\href@noop {} {\bibfield  {journal} {\bibinfo  {journal}
  {Nachr. Ges. Wiss. G{\"o}ttingen, Math.-Phys. Kl.}\ }\textbf {\bibinfo
  {volume} {3}},\ \bibinfo {pages} {213} (\bibinfo {year}
  {1904}{\natexlab{b}})}\BibitemShut {NoStop}%
\bibitem [{\citenamefont {Moore}(1920)}]{Moore_1920}%
  \BibitemOpen
  \bibfield  {author} {\bibinfo {author} {\bibfnamefont {E.~H.}\ \bibnamefont
  {Moore}},\ }\href@noop {} {\bibfield  {journal} {\bibinfo  {journal} {Bull.
  Amer. Math. Soc.}\ }\textbf {\bibinfo {volume} {26}},\ \bibinfo {pages} {394}
  (\bibinfo {year} {1920})}\BibitemShut {NoStop}%
\bibitem [{\citenamefont
  {Bjerhammar}(1951{\natexlab{a}})}]{Bjerhammar_1951_01}%
  \BibitemOpen
  \bibfield  {author} {\bibinfo {author} {\bibfnamefont {A.}~\bibnamefont
  {Bjerhammar}},\ }\href {https://doi.org/10.1007/BF02526278} {\bibfield
  {journal} {\bibinfo  {journal} {Bull. G{\'e}od{\'e}sique}\ }\textbf {\bibinfo
  {volume} {20}},\ \bibinfo {pages} {188} (\bibinfo {year}
  {1951}{\natexlab{a}})}\BibitemShut {NoStop}%
\bibitem [{\citenamefont
  {Bjerhammar}(1951{\natexlab{b}})}]{Bjerhammar_1951_02}%
  \BibitemOpen
  \bibfield  {author} {\bibinfo {author} {\bibfnamefont {A.}~\bibnamefont
  {Bjerhammar}},\ }\href@noop {} {\bibfield  {journal} {\bibinfo  {journal}
  {Trans. Roy. Inst. Tech. Stockholm}\ }\textbf {\bibinfo {volume} {49}}
  (\bibinfo {year} {1951}{\natexlab{b}})}\BibitemShut {NoStop}%
\bibitem [{\citenamefont {Bjerhammar}(1958)}]{Bjerhammar_1958}%
  \BibitemOpen
  \bibfield  {author} {\bibinfo {author} {\bibfnamefont {A.}~\bibnamefont
  {Bjerhammar}},\ }\href@noop {} {\bibfield  {journal} {\bibinfo  {journal}
  {Trans. Roy. Inst. Tech. Stockholm}\ }\textbf {\bibinfo {volume} {124}}
  (\bibinfo {year} {1958})}\BibitemShut {NoStop}%
\bibitem [{\citenamefont {Penrose}(1955)}]{Penrose_1955}%
  \BibitemOpen
  \bibfield  {author} {\bibinfo {author} {\bibfnamefont {R.}~\bibnamefont
  {Penrose}},\ }\href {https://doi.org/10.1017/S0305004100030401} {\bibfield
  {journal} {\bibinfo  {journal} {Proc. Cambridge Philos. Soc.}\ }\textbf
  {\bibinfo {volume} {51}},\ \bibinfo {pages} {406} (\bibinfo {year}
  {1955})}\BibitemShut {NoStop}%
\bibitem [{\citenamefont {Schr{\"o}dinger}(1930)}]{Schroedinger_1930}%
  \BibitemOpen
  \bibfield  {author} {\bibinfo {author} {\bibfnamefont {E.}~\bibnamefont
  {Schr{\"o}dinger}},\ }\href@noop {} {\bibfield  {journal} {\bibinfo
  {journal} {Sitz.-Ber. Preuss. Akad. Wiss., Phys.-Math. Kl.}\ }\textbf
  {\bibinfo {volume} {14}},\ \bibinfo {pages} {296} (\bibinfo {year}
  {1930})}\BibitemShut {NoStop}%
\bibitem [{\citenamefont {Holevo}(1982)}]{Holevo_1982}%
  \BibitemOpen
  \bibfield  {author} {\bibinfo {author} {\bibfnamefont {A.~S.}\ \bibnamefont
  {Holevo}},\ }\href@noop {} {\emph {\bibinfo {title} {Probabilistic and
  statistical aspects of quantum theory}}}\ (\bibinfo  {publisher}
  {North-Holland},\ \bibinfo {address} {Amsterdam},\ \bibinfo {year}
  {1982})\BibitemShut {NoStop}%
\bibitem [{\citenamefont {Lee}\ and\ \citenamefont
  {Tsutsui}(2020{\natexlab{c}})}]{Lee_2020_02}%
  \BibitemOpen
  \bibfield  {author} {\bibinfo {author} {\bibfnamefont {J.}~\bibnamefont
  {Lee}}\ and\ \bibinfo {author} {\bibfnamefont {I.}~\bibnamefont {Tsutsui}},\
  }\href@noop {} {\bibfield  {journal} {\bibinfo  {journal} {arXiv:2004.06099}\
  } (\bibinfo {year} {2020}{\natexlab{c}})}\BibitemShut {NoStop}%
\end{thebibliography}%

\end{document}